**Role of distal enhancers in shaping 3D-folding patterns and defining human-specific features of interphase chromatin architecture in embryonic stem cells**

Gennadi V. Glinsky[1]

[1] Institute of Engineering in Medicine

University of California, San Diego

9500 Gilman Dr. MC 0435

La Jolla, CA 92093-0435, USA

Correspondence: gglinskii@ucsd.edu

Web: http://iem.ucsd.edu/people/profiles/guennadi-v-glinskii.html

**Running title:** Evolution of human-specific regulatory networks

**List of abbreviations**

5hmC, 5-Hydromethylcytosine

CTCF, CCCTC-binding factor

DHS, DNase hypersensitivity sites

FHSRR, fixed human-specific regulatory regions

GRNs, genomic regulatory networks

HAR, human accelerated regions

hCONDEL, human-specific conserved deletions

hESC, human embryonic stem cells

HSGRL, human-specific genomic regulatory loci

HSNBS, human-specific NANOG-binding sites

HSTFBS, human-specific transcription factor-binding sites

LAD, lamina-associated domain

LINE, long interspersed nuclear element

lncRNA, long non-coding RNA

LTR, long terminal repeat

MADE, methylation-associated DNA editing

mC, methylcytosine

mESC, mouse embryonic stem cells

NANOG, Nanog homeobox

nt, nucleotide

POU5F1, POU class 5 homeobox 1

TAD, topologically associating domains

TE, transposable elements

TF, transcription factor

SE, super-enhancers

SED, super-enhancer domains

## Abstract

Molecular and genetic definitions of human-specific changes to genomic regulatory networks (GRNs) contributing to development of unique to human phenotypes remain a highly significant challenge. Genome-wide proximity placement analysis of diverse families of human-specific genomic regulatory loci (HSGRL) identified topologically-associating domains (TADs) that are significantly enriched for HSGRL and designated rapidly-evolving in humans TADs (Genome Biol Evol. 2016 8; 2774-88). Here, the analysis of HSGRL, hESC-enriched enhancers, super-enhancers (SEs), and specific sub-TAD structures termed super-enhancer domains (SEDs) has been performed. In the hESC genome, 331 of 504 (66%) of SED-harboring TADs contain HSGRL and 68% of SEDs co-localize with HSGRL, suggesting that emergence of HSGRL may have rewired SED-associated GRNs within specific TADs by inserting novel and/or erasing existing non-coding regulatory sequences. Consequently, markedly distinct features of the principal regulatory structures of interphase chromatin evolved in the hESC genome compared to mouse: the SED quantity is 3-fold higher and the median SED size is significantly larger. Concomitantly, the overall TAD quantity is increased by 42% while the median TAD size is significantly decreased ($p = 9.11E-37$) in the hESC genome. Present analyses illustrate a putative global role for HSGRL in shaping the human-specific features of the interphase chromatin organization and functions, which are facilitated by accelerated creation of new enhancers associated with targeted placement of HSGRL at defined genomic coordinates. A trend toward the convergence of TAD and SED architectures of interphase chromatin in the hESC genome may reflect changes of 3D-folding patterns of linear chromatin fibers designed to enhance both regulatory complexity and functional precision of GRNs by creating predominantly a single gene per regulatory domain structures.

## Introduction

Despite the remarkable progress illustrated by the near-comprehensive catalogue of changes within protein-coding genes that occurred during human evolution, the genetic and molecular basis defining unique to human phenotypic features remains largely unknown. These studies suggested that changes of protein-coding genes alone cannot explain most of the unique to human traits (Chimpanzee Sequencing and Analysis Consortium, 2005; Green et al., 2010; Meyer et al., 2012; Prüfer et al., 2012; 2014; Fu et al., 2014). Therefore, the hypothesis that unique to human phenotypes might result from human-specific changes to genomic regulatory sequences continues to remain highly relevant (King and Wilson, 1975).

In recent years, the extensive search for human-specific genomic regulatory loci (HSGRL) has resulted in identification of thousands candidate HSGRL, a vast majority of which is located within non-protein coding regions of human genome (McLean et al., 2011; Shulha et al., 2012; Konopka et al., 2012; Capra et al., 2013; Marnetto et al., 2014; Glinsky, 2015). The rapidly-evolving DNA sequences in humans have been identified that are highly conserved across mammals and have acquired many sequence changes in humans since divergence from chimpanzees (Pollard et al. 2006; Prabhakar et al. 2006; Bird et al. 2007). Subsequent experimental analyses of these sequences, which were designated human accelerated regions (HARs), revealed that some HARs function as non-coding RNA genes expressed during the neocortex development (Pollard et al. 2006) and human-specific developmental enhancers (Prabhakar et al. 2008). Recent computational and bioinformatics analyses and transgenic mouse experiments demonstrated that many HARs represent developmental enhancers (Capra et al., 2013).

Genome-wide analysis of transcription factor binding sites (TFBS) in human embryonic stem cells (hESC) demonstrated that transposable element-derived sequences, most notably LTR7/HERVH, LTR5_Hs/HERVK, and L1HS, harbor thousands of candidate human-specific TFBS (Glinsky, 2015). A majority of these candidate HSGRL appears to function as TFBS for NANOG, POU5F1 (OCT4), and CTCF proteins exclusively in hESC, suggesting their critical regulatory role during the early-stage embryogenesis (Glinsky, 2015). Interestingly, hESC-specific NANOG-binding sites appear enriched near the protein-coding genes regulating brain size, pluripotency long non-coding RNAs, hESC enhancers, and 5-hydroxymethylcytosine-harboring sequences immediately adjacent to TFBS. Consistent with the hypothesis that they contribute to development of human-specific phenotypes, candidate human-specific TFBS are placed near the coding genes associated with physiological development and functions of nervous and cardiovascular systems, embryonic development, behavior, as well as development of a diverse spectrum of pathological conditions such as cancer, diseases of cardiovascular and reproductive systems, metabolic diseases, multiple neurological and psychological disorders (Glinsky, 2015-2016). Taken together, these studies lend credence to the hypothesis that HSGRL may function in human cells as regulatory DNA sequences.

Significantly, a constantly expanding catalog of candidate HSGRL includes nearly twenty thousand of genomic regulatory elements comprising the following multiple distinct families of HSGRL:

i) regions of human-specific loss of conserved regulatory DNA termed hCONDEL (McLean et al., 2011);

ii) human-specific epigenetic regulatory marks consisting of H3K4me3 histone methylation signatures at transcription start sites in prefrontal neurons (Shulha et al., 2012);

iii) human-specific transcriptional genetic networks in the frontal lobe (Konopka et al., 2012);

iv) conserved in humans novel regulatory DNA sequences designated human accelerated regions, HARs (Capra et al., 2013);

v) fixed human-specific regulatory regions, FHSRR (Marnetto et al., 2014);

vi) human-specific transcription factor-binding sites, HSTFBS (Glinsky, 2015);

vii) DNase I hypersensitive sites (DHSs) that are conserved in non-human primates but accelerated in the human lineage (haDHS; Gittelman et al. 2015);

viii) DNase I hypersensitive sites (DHSs) that are under accelerated evolution, ace-DHSs (Dong et al., 2016).

In the mammalian nucleus, beads on a string linear strands of interphase chromatin fibers are folded into continuous megabase-sized structures termed topologically associating domains (TADs) that are readily detectable by the high-throughput analysis of interactions of chemically cross-linked chromatin (Dixon et al., 2012; Hou et al., 2012; Nora et al., 2012; Sexton et al., 2012). It has been suggested that TADs represent evolutionary-conserved spatially-segregated neighborhoods of high local frequency of intrachromosomal contacts reflecting the likelihood of individual physical interactions between long-range enhancers and promoters of target genes in live cells (Dixon et al., 2012; Gorkin et al., 2014). Formal definition of TADs implies that neighboring TADs are separated by the sharp boundaries, across which the intrachromosomal contacts are relatively infrequent (Dixon et al., 2012; Gorkin et al., 2014). A genome-wide proximity placement analysis of 18,364 DNA sequences representing multiple diverse families of HSGRL within the context of the principal regulatory structures of the interphase chromatin identified a sub-set of TADs that are significantly enriched for HSGRL and termed rapidly-evolving in humans TADs (Glinsky, 2016). There are significant enrichment within the boundaries of rapidly-evolving in humans TADs (revTAD) of hESC enhancers, primate-specific CTCF-binding sites, human-specific RNAPII-binding sites, hCONDELs, and H3K4me3 peaks with human-specific enrichment at transcription start sites (TSS) in prefrontal cortex neurons.

The identity of embryonic stem cells (ESC) critically depends on continuing actions of key master transcription factors (Young, 2011; Ng and Surani, 2011), which govern the maintenance of the ESC's pluripotency state by forming constitutively active super-enhancers (Hnisz et al., 2013; Whyte et al., 2013; Dowen et al., 2014). Recent experiments show that regulation of pluripotency in ESC occurs within selected TADs by establishing specific sub-TAD structures designated super-enhancer domains (Dowen et al., 2014). Super-enhancer domains (SEDs) are formed by the looping interactions between two CTCF-binding sites co-occupied by cohesin. In ESC, TADs and SEDs are designed to isolate super-enhancers (SEs) and target genes within insulated genomic neighborhoods to facilitate the functional precision of regulatory interactions between key genetic elements: SE, SE-driven cell identity genes and repressed genes encoding lineage-specifying developmental regulators (Dowen et al., 2014). Precise nature of structural-functional relationships between the HSGRL and principal structural elements of the interphase chromatin defining a spatial architecture of the pluripotency regulatory network in hESC remains unexplored.

Results of the present analyses revealed novel insights into structural-functional features of HSGRL of potential mechanistic significance that may have contributed to the marked changes of the interphase chromatin structure in the human genome, which are exemplified by the increased number and reduced size of TADs. Importantly, the concomitant HSGRL-associated increase of both the quantity and size of SEDs was documented in the hESC genome. Collectively, these observations identify the increasing of both regulatory complexity and functional precision of genomic regulatory networks due to convergence of TAD and SED architectures as one of the main trend of the interphase chromatin structural changes driven by the HSGRL during the evolution of *H. sapiens*.

## Results

### Markedly distinct features of super-enhancers in the hESC genome compared with mESC

The sustained expression of key cell identity genes and repression of genes encoding lineage-specifying developmental regulators is essential for maintaining ESC identity and pluripotency state. These processes are governed by the master TFs OCT4 (POU5F1), SOX2, and NANOG (OSN), that function by establishing SEs regulating cell identity genes, including master TFs themselves ((Hnisz et al., 2013; Whyte et al., 2013). The spatial relationships between SEs and HSGRL in the hESC genome within the context of TADs have not been investigated. To this end, the placement enrichment analysis was carried out to identify all TADs in hESC genome that harbor SEs and examine the association of SEs and HSGRL. There are 504 TADs (16%) harboring 642 SEs (94%) in the hESC genome. Remarkably, significant placement co-enrichments were observed between SEs, HSTFBS, and HARs residing within the boundaries of 279 TADs in the hESC genome (**Table 1**). In total, 279 TADs (8.9%) harbor 369 SEs (57.5%) that are co-localized with 300 HSTFBS and 564 HARs. When other HSGRL families (Glinsky, 2016) were considered in co-localization analyses, 331 TADs (10.6%) harboring 436 SEs (67.9%) were found to co-localize with HSGRL.

In human genome, there are approximately 7,000 high-confidence hESC-enriched enhancers (Xie et al., 2013; Hnisz et al., 2013), less than 10% of which are defined as SEs (Hnisz et al., 2013). Placement of high-confidence hESC enhancers is significantly enriched within the revTADs (Glinsky, 2016), implying possible genome-wide associations of HSGRL and hESC-enriched enhancers. Indeed, such associations became apparent when genomic co-localization analyses were performed to assess numbers of HSGRL and hESC-enriched enhancers residing within the boundaries of TADs. These analyses revealed that there is a significant direct correlation between the numbers of hESC-enriched enhancers and HARs that are located within the same revTADs (**Fig. 1**). Genome-wide placement and/or retention of HARs appears enriched within hESC enhancer-harboring TADs and seems to favor TADs containing larger numbers of hESC-enriched enhancers (**Fig. 1**).

Co-localization of a majority of SEs with HSGRL suggest that HSGRL may affect the structural-functional features of SEs in the hESC genome. Consistent with this hypothesis, genomes of human and mouse ESC manifest markedly distinct features associated with SE structures and functions. Despite strikingly similar genome sizes and numbers of protein-coding genes, hESC genome contains 3-fold more SEs compared to mouse: there are 684 SEs in the genome of hESC (Hnisz et al., 2013) and 231 SEs in the genome of mESC (Whyte et al., 2013). In the genome of hESC, only twenty-five of mESC SEs (11%) are represented as conserved orthologous sequences having genomic architecture of SEs, suggesting that 89% of mESC SEs lost the SE features in humans. Furthermore, 96% of SEs in the genome of hESC acquired structural-functional features of SEs during evolution after Euarchonta and Glires split 88 million years ago. The median size of SEs in hESC appears significantly larger compared to the median SE size in mESC ((9,589 bp versus 8,667 bp, respectively; $p = 0.017$). Detailed size distribution analyses demonstrated that accumulation of SEs in hESC genome is associated with increased number of large SEs and decreased number of small SEs compared to mouse genome: there are 7-fold increase of very large SEs having size more than 30 Kb, consistent ~3-fold increase of SEs having size range from 2 – 30 Kb, and a marked 38-fold depletion of small SEs having size less than 2 Kb (**Table 2; Fig. 2**).

Collectively, these data indicate that structural-functional features of SEs are markedly distinct in the hESC genome compared to the mouse, which appear associated with the enrichment of HSGRL within SE-harboring TADs. It will be of interest to determine whether the HSGRL placement within SE-harboring TADs exerts biologically-meaningful effects on SE functions. Targeted placements and/or retention of HSGRL may increase density of enhancer elements within selected TADs, which would result in the merger of conventional enhancer units into super-enhancer structures containing the exceptionally high level of transcription-enabling apparatus to drive and continually maintain high expression of associated target genes. This idea is supported by the findings that nearly 60% (1571) of the HARs overlap at least one of the common markers of enhancer activity in human cells (Capra et al., 2013).

**TAD structural features are markedly altered in the hESC genome compared with mESC**

There are 3,127 TADs in the hESC genome, which is 42% more than 2,200 TADs in the genome of mESC (Dixon et al., 2012). Correspondingly, there are 87,883 CTCF-binding sites in hESC (Kunarso et al., 2010), 29,018 (33%) of which represent primate-specific CTCF-binding sequences (Glinsky, 2015). The median size of TADs in hESC is significantly smaller compared to the median size of TADs in mESC (680 Kb versus 880 Kb; p = 9.11E-37). Detailed size distribution analyses of TADs in human and mouse ESC (**Fig. 2; Table 3**) revealed that there is a ~ 2-fold depletion of large TADs having size > 2,000 Kb and consistently increased numbers of medium-size and small TADs having size range from 100 - 1,000 Kb (**Table 3**). Structural-functional features of the revTADs in the hESC genome appear markedly distinct from the TADs of the orthologous sequences in the mouse ESC genome. Of the 60 revTAD (Glinsky, 2016), nineteen (32%) are placed within the primate-specific sequences that failed to align to the mouse reference genome database sequence. Remaining revTADs appear to evolve by erasing the existing and establishing new domain boundaries within orthologous DNA sequences via two distinct mechanisms: 1) domain & boundary crossing (hESC TADs appear to cross boundaries of 2 to 4 orthologous TADs in mESC genome); and 2) domain shrinking & boundary creation (smaller hESC TADs are placed within the boundaries of larger orthologous TAD sequences in mESC genome).

The above considerations prompted additional analyses of relationships between the SEs and TADs in the genomes of human and mouse ESC (**Fig. 3**). These analyses revealed that in hESC there are significant direct correlations between the size of TADs and SE's span defined as a number of bp between the two most distant SEs located within a given TAD (**Fig. 3A**). Similar trends were observed in the mESC genome, however, the correlation coefficient values were not statistically significant. Correlation patterns observed for the 60 revTADs for placements of hESC-enriched enhancers, HARs, HSTFBS, and size of TADs were validated on a larger set of 147 revTADs (**Fig. 3B**). Genome-wide, the highly significant direct correlation was discovered between the size of TADs and the number of hESC-enriched enhancers located within TADs (**Fig. 3C**). These observations are conceptually coherent because TAD boundaries were inferred from the relative prevalence and directionality of interchromosomal interactions along the chromosome length (Dixon et al., 2012), which are predominantly mediated by the enhancers' activities and detected as enhancer-promoter and enhancer-enhancer interactions in the Hi-C analyses. It will be of interest to determine experimentally whether the size and boundaries of the adjacent and neighboring TADs can be altered by increasing the density, structure, and activity of the resident enhancers.

**Potential mechanisms of HSGRL-mediated effects on principal regulatory structures of interphase chromatin**

Present analyses provide a conceptual framework for understanding genome-scale regulatory changes during evolution within the context of the principal regulatory structures of the interphase chromatin to reflect an apparent trend toward increasing complexity of genomic regulatory networks (**Fig. 4**). Experimental observations at the foundation of building blocks of the genome's evolution model are described in the previous sections and additional considerations are focused on the analyses of potential contributions of hESC-enriched enhancers, SEs, and SEDs to these processes.

According to the model, one of the key elements of the evolution of genomic regulatory networks is the creation of new enhancer elements (**Fig. 4**). Conventional enhancers comprise discrete DNA segments occupying a few hundred base pairs of the linear DNA sequence and harboring multiple TFBS. SEs consist of clusters of conventional enhancers that are densely occupied by the master transcription factors and Mediator (Whyte et al., 2013). Therefore, it is logical to expect that creation of new TFBS and increasing density of TFBS would increase the probability of the emergence of new enhancer elements. It follows that creation of new enhancers and increasing their density would facilitate the emergence of new SE structures. This sequence of events would imply that evolutionary time periods required for creation of TFBS, enhancers, and SEs are shortest for TFBS, intermediate for enhancers, and longest for SEs. To test this assumption, estimates of creation time periods for enhancers and SEs in the hESC genome were calculated (**Table 4**) and compared to the previously reported estimates of creation time periods of TFBS (Glinsky, 2015). Consistent with the model expectations, the estimated creation time is markedly longer for SEs compared to enhancers for conserved (8-fold), primate-specific (17-fold), and human-specific (63-fold) regulatory sequences (**Table 4**). Furthermore, the estimated creation time periods for enhancers appear several fold longer compared to the creation time estimates for TFBS sequences in both chimpanzee and humans (Glinsky, 2015). One of the intriguing results of this analysis is that 87% of new SEs and 75% of new enhancers in the hESC genome were created within conserved sequences (**Table 4**). Similarly, a majority (67%) of new binding sites for NANOG (59,221 of 88,351 TFBS) and CTCF (58,865 of 87,883 TFBS) proteins in the hESC genome compared with mESC are located within conserved sequences (**Supplemental Fig. S3**).

Notably, the estimated creation time appears accelerated in humans compared to chimpanzee for both SEs (7-fold) and enhancers (27-fold), suggesting that human genomes were acquiring new regulatory elements and increasing the regulatory complexity and precision of genomic regulatory networks at the markedly accelerated pace.

CTCF-binding sites play a crucial role in defining the TAD boundaries (Dixon et al., 2012; Li et al., 2013) and in establishing the SED architecture (Dowen et al., 2014). Knocking down the expression of CTCF and cohesin genes to reduce their protein concentrations results in a decrease in the number of intra-TAD chromatin loops (Sofueva et al., 2013; Seitan et al., 2013; Zuin et al., 2014), increase in the number of inter-TAD chromatin interactions crossing the TAD boundaries (Zuin et al., 2014), and chromatin compaction (Tark-Dame et al., 2014). Several lines of evidence are in agreement with the hypothesis that creation of new CTCF-binding sites may have contributed to the rewiring of interphase chromatin architecture and genomic regulatory networks during primate evolution:

i) In the hESC genome, 29,018 of 87,883 (33%) CTCF-binding sites represent primate-specific sequences (Kunarso et al., 2010; Glinsky, 2015);

ii) Of the 23,709 constitutive CTCF-binding sites implicated in defining TAD boundaries in the human genome (Li et al., 2013), 6,787 sites (28.6%) failed alignment to the mouse genome, suggesting that they represent primate-specific sequences;

iii) Two hundred eighty seven HARs (10.5%) are located within 10 Kb of the 336 constitutive CTCF-binding sites;

iv) Five hundred seventy one HARs (20.8%) are located within 1 Kb of 953 CTCF-binding sites;

v) Activation of retrotransposons and lineage-specific repeat-driven dispersion of CTCF-binding sites has produced species-specific expansions of CTCF binding in mammalian genomes and these new

CTCF-binding sites function as chromatin domain insulators and transcriptional regulators (Schmidt et al., 2012).

Detailed proximity placement and HARs/TFBS co-localization analyses identified 123 HARs located within 1 Kb from 127 high-confidence CTCF-binding sites, which were validated in 23 different human cells lines. Notably, 123 of 127 (97%) high-confidence CTCF-binding sites represent overlapping CTCF/RAD21-binding sites, which is consistent with their putative regulatory role in establishing TAD boundaries and/or SED architecture. A significant majority of 123 HARs located near high-confidence overlapping CTCF/RAD21-binding sites harbor at least one TFBS in human cells (76 HARs; 62%).

Significant increase of TAD and SED numbers in human genomes compared with mouse indicates that chromatin folding profiles are dramatically altered, reflecting more frequent intrachromosomal segmentation of linear DNA fibers and enhanced likelihoods of formation of isolated intra-segmental looping structures. These changes of chromatin folding profiles and DNA fibers' packaging patterns would impose more stringent requirement on DNA elasticity, particularly, near the TAD and SED boundaries due to bending of DNA double helix. Consequently, changes of DNA sequence and/or chromatin structure may be necessary to accommodate these new structural requirements. It has been reported that SINE repeats are enriched at the TAD boundaries (Dixon et al., 2012), suggesting that insertion of repetitive elements may contribute to changes of DNA elasticity near putative bending sites. Consistent with this hypothesis, a survey of SED sequences in hESC revealed an apparent systematic placement of Alu elements within ~5 Kb windows near SED boundaries (**Fig. 5**). Clusters of closely-spaced sequences of at least three Alu elements belonging to most ancient AluJ (~65 million years old), second oldest AluS (~30 million years old), and currently active modern AluY sub-families (Bennett et al., 2008) were observed frequently, suggesting that placement and/or retention of Alu elements at these sites were occurring for millions of years and continues at the present time.

Taken together with the recent report that nearly 60% of the HARs overlap at least one of the common markers of enhancers in human cells (Capra et al., 2013), these observations strongly argue that creation of new enhancer elements is one of the key events defining evolution of genomic regulatory networks. Marked acceleration of the TFBS and enhancers' creation processes in humans may have contributed to genome-scale rearrangements of principal regulatory structures of the interphase chromatin, leading to the increased complexity and enhanced precision of genomic regulatory networks and emergence of human-specific phenotypes. Frequent locations of Alu sequences near the putative DNA bending sites suggest their potential role in regulation of the elasticity of chromatin fibers by reducing nucleosome placements due to the low affinity of Alu elements to nucleosome binding (Huda et al., 2009) and/or facilitating strand invasion reactions between neighboring DNA strands leading to Alu/Alu recombination events (Lee at al., 2015; Morales et al., 2015). This model is in agreement with the observations that Alu elements appear preferentially retained in GC-rich and gene-rich regions of the human genome (Lander et al., 2001).

**Conservation patterns of HSGRL in individual human genomes**

All analyses conducted so far were performed using the reference genome databases and not the individual human genomes. To address this limitation, the assessment of conservation of HSGRL in individual genomes of 3 Neanderthals, 12 Modern Humans, and the 41,000-year old Denisovan genome (Reich et al., 2010; Meyer et al., 2012) was carried-out by direct comparisons of corresponding sequences retrieved from individual genomes and the human genome reference database (http://genome.ucsc.edu/Neandertal/ ). Full-length sequence alignments with no gaps of the individual genome sequences to the corresponding sequences in the human genome reference databases of both hg18 and hg19 releases were accepted as the evidence of sequence conservation in the individual human genome.

Sequences of all analyzed to date HSGRL appear conserved in individual human genomes (**Fig. 6** and **Supplemental Fig. S4**), albeit a significant inter-individual variability in degree of conservations of specific HSGRL is apparent. Conservation of all HSGRL are consistently at the lowest level in the Neanderthals' genome, suggesting that creation and/or retention rates of HSGRL are enhanced in Modern Humans. The results of these analyses indicate that sequences of HSGRL with assigned biochemical functions, e.g., specific TFBS or Lamin B1 (LMNB1)-binding sites, which are residing within HARs, exhibit markedly higher conservation levels compared to sequences of HARs harboring the corresponding HSGRL. This conclusion remains valid for HSGRL sequences with assigned specific biochemical or biological functions that were associated with HARs by proximity placement analyses (**Figs. 6A, B**). HSGRL sequences manifesting the relatively high conservation levels in the Neanderthals' genome appear most conserved in the individual genomes of Modern Humans as well, including the 41,000-year old Denisovan genome (**Fig. 6B**).

Consistent patterns of significant direct correlations between conservation profiles of distinct seemingly unrelated HSGRL sequences in individual human genomes were observed (**Fig. 6C** and **Supplemental Fig. S4**). One possible interpretation of these observations is that HSGRL conservation patterns reflect intrinsic features of the individual human genome, integration of which can gauge the overall capacity of a genome to create and retain HSGRL. This idea was tested by calculating for each individual human genome the genomic fitness scores integrating into a single numerical value sequence conservation data of 909 HSGRL (**Fig. 6C;** bottom panel). Notably, individual human genomes appear markedly distinct based on the results of this analysis with the lowest genomic fitness score of 0.9 in Neanderthals, followed by the scores of 1.89; 2.22; and 2.27 for individual genomes of Native American, Denisova cave, and Mongolian subjects, respectively. On the other end of the spectrum, the genomic fitness scores of 4.51; 4.72; and 4.73 were obtained for genomes of Yoruba (West Africa), French (Western Europe), and San (Southern Africa) individuals, respectively.

These observations are highly congruent with the recent definition of structural variations-free (fixed) human-specific regulatory regions (FHSRR) based on the stringent formal analysis of the HSGRL intra-species patterns of variations within human population using exome and full genome sequencing database of 1,092 individuals from the 1000 Genomes Project Consortium (Abecasis et al., 2012; Marnetto et al., 2014). Collectively, the consistent evidence of HSGRL conservation in individual human genomes are in accord with the hypothesis of their putative functional role in defining human-specific phenotypes.

## Discussion

Placing several meters of linear DNA fibers into micron-sized nucleus while maintaining the genome integrity and ensuring the functional precision of genomic regulatory networks represent a set of formidable challenges successfully addressed during the evolution of eukaryotes. The elegant evolutionary solutions to these problems are illustrated by the recent pioneering work on the 3D structures of interphase chromosomes. These studies revealed specific folding patterns of the linear chromosome fibers into spatially-segregated domain-like segments experimentally-defined as the continuous megabase-sized topologically associating domains (TADs). TADs are readily and reproducibly detectable by the high-throughput analysis of inter-chromosomal interactions of chemically cross-linked chromatin (Dixon et al., 2012; Hou et al., 2012; Nora et al., 2012; Sexton et al., 2012). It has been suggested that one of the principal creative events in the continuing chromatin domain architecture remodeling process during human evolution is the HSGRL-enabled emergence of new enhancer elements (Glinsky, 2016-2017), increasing density of which would enhance the probability of structural transition from conventional enhancers to novel SE structures and the subsequent formation of new SEDs (**Fig. 4**). Significantly, primate-specific and human-specific sequences appear to contribute to creation of interacting pairs of conserved and/or newly created overlapping CTCF/cohesin-binding sites flanking novel SEs, which represents the essential structural requirement for the formation of functional SEDs. On the evolutionary scale, these processes involve continuing removal of old and creation of new regulatory sequences through multiple trial-and-error events enabled by retrotransposition, cytosine MADE, and recombination (Glinsky, 2015-2016). Collectively, the ensemble of these HSGRL-associated changes of the nuclear regulatory architecture may facilitate the enhanced precision of regulatory interactions between enhancers and target genes within specifically reconstructed insulated SED neighborhoods and rewired genome-wide TAD networks. Taken together, these observations imply that the convergence of TAD and SED architectures, which is exemplified in the human genome by concomitant processes of increasing both quantity and size of SEDs and the increasing number and size reduction of TADs, might represent one of the main directions of the interphase chromatin structural changes during the evolution of genomic regulatory networks.

Recent studies documented several human-specific features of the interphase chromatin architecture, evolution of which appear associated with the targeted placement and/or retention of HSGRL. Genome-wide proximity placement analyses of 18,364 DNA sequences representing distinct families of HSGRL identified a small fraction of TADs in the human genome, which manifest a statistically significant accumulation of HSGRL compared to the expected values estimated based on a random distribution model (Glinsky, 2016). This set of TADs, termed rapidly-evolving in humans TADs (revTADs), acquired the maximum enrichment levels of 16-fold for HARs; 17-fold for hESC-FHSRRs; 50-fold for HSTFBS; and 88-fold for DHS-FHSRRs ($p < 0.0001$ in all instances; Glinsky, 2016). Follow-up analyses of sixty revTADs, which were defined based on the enrichment of HARs and HSTFBS, revealed that all reported to date HSGRL and human-specific epigenetic signatures associated with embryonic development appear significantly enriched within the revTAD boundaries (Glinsky, 2016). One of the features of revTADs was revealed by the correlation analyses of HARs and HSTFBS: placements of different types of HSTFBS manifest significant positive correlations, whereas placements of HARs and HSTFBS exhibit inverse correlation profiles (Glinsky, 2016). Consistent with the hypothesis of the increased precision of genomic regulatory networks during evolution, a majority of revTADs in human genome tend to harbor a small number of coding genes: 65% of revTADs contain five or less protein-coding genes, among which eight revTADs harbor just one protein-coding gene and two revTADs contain only non-coding RNA genes. Several revTADs contain large clusters of functionally-related protein-coding genes: three revTADs on chr11 harbor 61 genes encoding olfactory receptors; three consecutively-spaced revTADs on chr19 harbor 31 genes encoding zinc finger proteins; and one revTAD on chr20 contains eight beta-defensins' genes. Overview of the RNAseq data documented two other features of revTADs: i) clearly discernable tissue-specific patterns of non-coding RNA expression; and ii) a prominent presence of non-coding RNA transcripts in human brain regions.

Correlation screens of proximity placement patterns of different HSGRL residing within the revTADs revealed markedly distinct correlation profiles of individual members of HSGRL families and recombination rates within the host revTADs. These analyses readily distinguish placement patterns of HARs and HSTFBS: revTADs having high recombination scores tend to accumulate large numbers of HARs while revTADs with low recombination scores harbor high numbers of HSTFBS (Glinsky, 2016). Consistent with the requirements of DNA strand interactions and strand invasion for the recombination process, significant direct correlations were observed between recombination scores and intra-chromosomal contacts within revTADs (Glinsky, 2016). Therefore, placements of HARs and HSTFBS within revTADs appear associated with distinct molecular processes. Placement of HARs within revTADs exhibits significant positive correlation with high recombination rates, which seems to connect the biogenesis of HARs with recombination mechanisms. In contrast, significant inverse correlation between HSTFBS placement and recombination rates as well as location of 99% of HSTFBS within TE – derived DNA sequences (Glinsky, 2015) strongly implicate TE activity in the biogenesis of HSTFBS. Since HARs by definition are located within highly evolutionary conserved DNA sequences, results of these analyses suggest that molecular mechanisms driving the emergence of regulatory loci evolved by exaptation of ancestral DNA (Cotney et al., 2013; Villar et al., 2015) may be associated with meiotic recombination as well.

One of the notable features of potential mechanistic significance is the association of hESC-enriched enhancers and SEDs with HSGRL (**Table 1**). Taking into account that there are ~4,000 HSTFBS in the hESC genome (Glinsky, 2015) and nearly 60% of HARs overlap at least one of the common markers of enhancers in human cells (Capra et al., 2013), it seems logical to propose that creation of new enhancer elements leading to increasing density of conventional enhancers in selected genomic regions is one of the key events defining the increasing genomic complexity as the main direction of evolution of GRNs. Consistent with this idea, there are significant direct correlations between the placements within TADs of HARs and hESC-enriched enhancers (**Fig. 1**), indicating that there is an apparent trend of the placement preference of HARs within TADs harboring hESC-enriched enhances. It follows, that higher density of conventional enhancers would increase the probability of local structural transitions of chromatin architecture to SEs and SEDs, provided other local structural requirements are in place or co-evolved (see below). In agreement with this model, there are 3-fold more SEs in the hESC genome compared with the mESC and the median size of hESC SEs is significantly larger (**Fig. 2**).

Concomitantly, marked changes of TAD structural features in the hESC genome compared with mESC were observed, which are particularly striking for revTADs. Genome-wide, there are 42% more TADs in the hESC genome and the median size of hESC TADs is significantly smaller (**Fig. 2**). These changes of TAD structural features remain consistent when the corresponding comparisons were made between the similar in size individual human and mouse chromosomes (**Fig. 2**), indicating that hESC genome contains the increased number of the predominantly smaller size TADs. Consistent with these observations, comparisons of the high-resolution 3D maps of human and mouse genomes, which were recently obtained by Rao et al. (2014) using in situ Hi-C at a resolution range of 1 – 5 Kb, revealed that there are ~3-fold more contact domains formed by the concomitantly increased numbers of long-range chromatin loops in the human GM12878 B-cell lymphoblasts compared with the mouse CH12 B-cell lymphoblasts (Rao et al., 2014).

In the hESC genome, structural changes of the revTADs seem particularly evident: 32% of revTADs are located within primate-specific genomic regions and the remaining revTADs appear evolved via mechanisms of boundary crossing, domain mergers, and creation of new boundaries within larger TADs of orthologous mouse sequences. Collectively, this dramatic changes of the regulatory infrastructure of interphase chromatin can be explained by the model of convergence of TAD and SED architecture (**Fig. 4**). According to this model, the high density of conventional enhancers increases the likelihood of transition to larger in size SE structures, formation of new SEDs, and increasing segmentation of genomic regions into insulated regulatory neighborhoods of large SED/small TAD structures. Consistent with this model, significant correlations were observed between the size of TADs and genomic span of hESC SEs residing within TADs (**Fig. 3A**). There are significant direct correlations between the revTAD sizes and the numbers of hESC-enriched enhancers residing within revTADs (**Fig. 3B**). Genome-wide, the highly significant direct correlation was observed between the size of TADs and the number of hESC-enriched enhancers located within TADs (**Fig. 3C**).

In addition to enabling the creation of new enhancer elements, increasing density of conventional enhancers, and facilitating transition to SED structures, potential mechanisms of HSGRL-mediated effects on principal regulatory structures of interphase chromatin are likely involve emergence of overlapping CTCF/cohesin sites and LMNB1-binding sites as well as continuing insertion of Alu clusters near the putative DNA bending sites (**Figs. 4 & 5**). Collectively, the ensemble of these structural changes facilitated by the targeted placements and retention of HSGRL at specific genomic locations would enable the emergence of new SED structures and remodeling of existing TADs to drive evolution of GRNs (**Fig. 4**). Presented models of dynamic transitions of the interphase chromatin principal regulatory structures are in accord with the results of recent high-resolution in situ Hi-C experiments demonstrating that human genomes are partitioned into contact domains consisting of ~10,000 loops accommodating functional links between enhancers and promoters (Rao et al., 2014). Consistent with the idea that chromatin loops frequently demarcate the boundaries of contact domains, anchors at the loop bases typically occur at the contact domain boundaries and involve binding of two CTCF/cohesin sites in a convergent orientation with the asymmetric binding motifs of interacting sites aligned to face each other (Rao et al., 2014).

Compelling evidence of conservation in individual human genomes of different families of HSGRL (Marnetto et al., 2014; Glinsky, 2015; 2016 **Fig. 6; Supplemental Fig. S4**) are in accord with the hypothesis of their putative functional role in defining human-specific phenotypes. Observations of highly consistent patterns of significant direct correlations between conservation profiles of distinct seemingly unrelated HSGRL sequences in individual human genomes were observed (**Fig. 6C** and **Supplemental Fig. S4**) suggest that HSGRL conservation patterns reflect intrinsic features of the individual human genome, integration of which can help to assess the overall capacity of a genome to create and retain HSGRL. In light of emerging evidence highlighting the role of HSGRL in development of therapy-resistant malignancies (Glinsky, 2015-2016), it will be of interest to explore experimentally the potential practical utility of the outlined herein concepts by building and validating the working models of defined super-enhancers' domains and associated TAD structures in the hESC genome (Glinsky, 2017).

**Concluding remarks**

In general terms, chromosomes can be viewed as the physical conduits of genetic information enabling its secure storage, maintenance, transfer, and efficient translation into a diverse spectrum of specific phenotypes. To a large degree all these processes are facilitated by the linear code of DNA sequences and enabled by the chromatin structures defining a 3D architecture of interphase chromosomes, which can be considered collectively as the chromatin folding code. Present analyses suggest that increasing regulatory complexity in human genomes associated with targeted placements of many thousands of HSGRL has a major effect on the principal regulatory structures of interphase chromatin, namely TADs, SEs, and SEDs, perhaps, contributing to the creation of human-specific chromatin folding code. Recently reported half-lives and mean-lifetimes of enhancers were estimated at 296 and 427 hundred million years, respectively (Villar et al., 2015). Comparisons of these estimates with the estimates of time periods required for creation of enhancers and SEs (**Table 4**) seem to indicate that new enhancer elements are created at a markedly faster pace compared with their decay time span. This marked evolutionary dichotomy of time period requirements for creation of new enhancers compared to the enhancers' loss provides an underlying mechanism explaining why the increasing genomic complexity represents a major trend during the evolution of genomic regulatory networks.

Exaptation of ancestral DNA was identified as a main mechanism of creation of human-specific enhancers active in embryonic limb (Cotney et al., 2013) and as a prevalent creation mechanism of recently evolved enhancers during the mammalian genome evolution (Villar et al., 2015). Consistently, a vast majority of distinct classes of regulatory elements in the hESC genome appears created on conserved DNA sequences (**Table 4**; **Supplemental Fig. S3**), suggesting that exaptation of ancestral DNA constitutes a main mechanism of creation of new regulatory sequences in the human genome. Significantly, exaptation of ancestral DNA appears to generate overall many more recently evolved regulatory sequences than can be attributed to the repeat-driven expansion mechanisms. Therefore, these observations provide further support to the hypothesis that meiotic recombination is the predominant mechanism responsible for creation of new regulatory loci during evolution.

The concept of evolution of evolvability posits that a genotypic feature evolves not due to its functional effect, but due to its effect on the ability of DNA to evolve more quickly (Wagner and Altenberg, 1996). Considering the evolution of enhancers as an example of evolution of evolvability (He et al., 2012; Duque et al., 2014) and taking into account the potential role of HSGRL in creation of human-specific networks of enhancers, SEs, and SEDs, the apparent mechanistic links of these processes to several key enzymatic systems should be regarded as highly promising novel molecular targets of potential therapeutic significance. Of particular relevance in this context is the idea that the regulatory balance of HAT and HDAC activities at specific genomic loci as well as the performance of enzymatic systems regulating cytosine recovery/methyl-cytosine deamination cycles and DNA recombination processes may play a major role during the evolution of regulatory DNA sequences and emergence of genomic regulatory networks controlling unique to human physiological and pathological phenotypes. Emerging evidence implicating HSGRL derived from stem cell-associated retroviral sequences in the pathogenesis of clinically-lethal human malignancies (Glinsky, 2015-2016) seem particularly intriguing in this regard.

## Materials and Methods

*Data Sources and Analytical Protocols*

Solely publicly available datasets and resources were used for this analysis as well as previously reported methodological approaches and computational pipelines validated for discovery of primate-specific gene and human-specific regulatory loci (Tay et al., 2009; Kent, 2002; Schwartz et al., 2003; Capra et al., 2013; Marnetto et al., 2014; Glinsky, 2015). The analysis is based on the University of California Santa Cruz (UCSC) LiftOver conversion of the coordinates of human blocks to corresponding non-human genomes using chain files of pre-computed whole-genome BLASTZ alignments with a minMatch of 0.95 (unless specified otherwise) and other search parameters in default setting (http://genome.ucsc.edu/cgi-bin/hgLiftOver). Extraction of BLASTZ alignments by the LiftOver algorithm for a human query generates a LiftOver output "Deleted in new", which indicates that a human sequence does not intersect with any chains in a given non-human genome. This indicates the absence of the query sequence in the subject genome and was used to infer the presence or absence of the human sequence in the non-human reference genome. Human-specific regulatory sequences were manually curated to validate their identities and genomic features using a BLAST algorithm and the latest releases of the corresponding reference genome databases for time periods between April, 2013 and June, 2015.

Genomic coordinates of 3,127 topologically-associating domains (TADs) in hESC; 6,823 hESC-enriched enhancers; 6,322 conventional and 684 super-enhancers (SEs) in hESC; 231 SEs and 197 SEDs in mESC were reported in the previously published contributions (Dixon et al., 2012; Xie et al., 2013; Hnisz et al., 2013; Whyte et al., 2013; Dowen et al., 2014). The primary inclusion criterion for selection of the human-specific genomic regulatory loci (HSGRL) analyzed in this contribution was the fact that they were identified in human cells lines and primary human tissues whose karyotype were defined as "normal". The following four HSGRL families comprising of 10,598 individual regulatory DNA sequences were analyzed in this study: 1) Human accelerated regions (HARs; Capra et al., 2013); 2) Human-specific transcription factor-binding sites (HSTFBS; Glinsky, 2015); 3) hESC-derived fixed human-specific regulatory regions (hESC-FHSRR; Marnetto et al., 2014); 4) DNase hypersensitive sites-derived fixed human-specific regulatory regions (DHS-FHSRR; Marnetto et al., 2014). The number of HSGRL placed within a given TAD was computed for every TAD in the hESC genome and the HSGRL placement enrichment was calculated as the ratio of observed values to expected values estimated from a random distribution model at the various cut-off thresholds. Datasets of NANOG-, POU5F1-, and CTCF-binding sites and human-specific TFBS in hESCs were reported previously (Kunarso et al., 2010; Glinsky, 2015) and are publicly available. RNA-Seq datasets were retrieved from the UCSC data repository site (http://genome.ucsc.edu/; Meyer et al., 2013) for visualization and analysis of cell type-specific transcriptional activity of defined genomic regions. A genome-wide map of the human methylome at single-base resolution was reported previously (Lister et al., 2009; 2013) and is publicly available (http://neomorph.salk.edu/human_methylome). The histone modification and transcription factor chromatin immunoprecipitation sequence (ChIP-Seq) datasets for visualization and analysis were obtained from the UCSC data repository site (http://genome.ucsc.edu/; Rosenbloom et al., 2013). Genomic coordinates of the RNA polymerase II (PII)-binding sites, determined by the chromatin integration analysis with paired end-tag sequencing (ChIA-PET) method, were obtained from the saturated libraries constructed for the MCF7 and K562 human cell lines (Li et al., 2012). Genome-wide maps of interactions with nuclear lamina, defining genomic coordinates of human and mouse lamin-associated domains (LADs), were obtained from previously published and publicly available sources (Guelen et al., 2008; Peric-Hupkes et al., 2010). The density of TF-binding to a given segment of chromosomes was estimated by quantifying the number of protein-specific binding events per 1-Mb and 1-kb consecutive segments of selected human chromosomes and plotting the resulting binding site density distributions for visualization. Visualization of multiple sequence alignments was performed using the WebLogo algorithm (http://weblogo.berkeley.edu/logo.cgi ). Consensus TF-binding site motif logos were previously reported (Kunarso et al., 2010; Wang et al., 2012; Ernst and Kellis, 2013).

The quantitative limits of proximity during the proximity placement analyses were defined based on several metrics. One of the metrics was defined using the genomic coordinates placing HSGRL closer to putative target protein-coding or lncRNA genes than experimentally defined distances to the nearest targets of 50% of the regulatory proteins analyzed in hESCs (Guttman et al., 2011). For each gene of interest, specific HSGRL were identified and tabulated with a genomic distance between HSGRL and a putative target gene that is smaller than the mean value of distances to the nearest target genes regulated by the protein-coding TFs in hESCs. The corresponding mean values for protein-coding and lncRNA target genes were calculated based on distances to the nearest target genes for TFs in hESC reported by Guttman et al. (2011). In addition, the proximity placement metrics were defined based on co-localization within the boundaries of the same TADs and the placement enrichment pattern of HSNBS located near the 251 neocortex/prefrontal cortex-associated genes, which identified the most significant enrichment of HSNBS placement at the genomic distances less than 1.5 Mb with a sharp peak of the enrichment p value at the distance of 1.5 Mb (**Supplemental Figure S7**).

The assessment of conservation of HSGRL in individual genomes of 3 Neanderthals, 12 Modern Humans, and the 41,000-year old Denisovan genome (Reich et al., 2010; Meyer et al., 2012) was carried-out by direct comparisons of corresponding sequences retrieved from individual genomes and the human genome reference database (http://genome.ucsc.edu/Neandertal/ ). Direct access to the specific Genome Browser tracks utilized for analyses and visualization:

http://genome.ucsc.edu/cgi-bin/hgTracks?db=hg18&position=chr10%3A69713986-69714099&hgsid=393865029_yg7UixUE4a4awjjTahns4KTPkIl1 .

Recombination rates were downloaded from the HapMap Project (The International Hapmap Consortium, 2007) and the numbers of DNA segments with the recombination rates of 10 cM/Mb or greater were counted. This threshold exceeds ~10-fold the mean intensity of recombination rates in telomeric regions, which were identified as the regions with the higher recombination rates in the human genome. It is well known that over large genomic scales, recombination rates tend to be higher in telomeric as compared to centromeric chromosomal regions. In telomeric regions, the mean detected hotspot spacing is 90 kb and the mean intensity (total rate across the hotspot) per hotspot is 0.115 cM, whereas for centromeric regions the mean spacing is 123 kb and the mean intensity is 0.070 cM (The International Hapmap Consortium, 2007).

*Statistical Analyses of the Publicly Available Datasets*

All statistical analyses of the publicly available genomic datasets, including error rate estimates, background and technical noise measurements and filtering, feature peak calling, feature selection, assignments of genomic coordinates to the corresponding builds of the reference human genome, and data visualization, were performed exactly as reported in the original publications and associated references linked to the corresponding data visualization tracks (http://genome.ucsc.edu/). Any modifications or new elements of statistical analyses are described in the corresponding sections of the Results. Statistical significance of the Pearson correlation coefficients was determined using GraphPad Prism version 6.00 software. The significance of the differences in the numbers of events between the groups was calculated using two-sided

Fisher's exact and Chi-square test, and the significance of the overlap between the events was determined using the hypergeometric distribution test (Tavazoie et al., 1999).

## Supplemental Information

Supplemental information includes Supplemental Figures S1–S4 and can be found with this article online.

## Author Contributions

This is a single author contribution. All elements of this work, including the conception of ideas, formulation, and development of concepts, execution of experiments, analysis of data, and writing of the paper, were performed by the author.

## Acknowledgements

This work was made possible by the open public access policies of major grant funding agencies and international genomic databases and the willingness of many investigators worldwide to share their primary research data. I would like to thank my colleagues for their valuable critical contributions during the informal review and formal peer review process of this work.

**Figure Legends**

**Figure 1.** Genome-wide associations between the number of HARs and the quantity of hESC-enriched enhancers located within TADs.

A. Direct correlation between the numbers of HARs and hESC-enriched enhancers located within the individual revTADs.

B-D. In the hESC genome, TADs segregated into sub-groups with increasing numbers of HARs contain concomitantly higher proportion of hESC-enriched enhancer-harboring TADs compared with TADs without hESC enhancers. The fractions of TADs with and without hESC enhancers are shown for a total of 3,062 TADs (hg19 release of human reference genome database) in the hESC genome (B), for 1,135 TADs harboring 2,609 HARs (C), for 2,075 TADs containing 6,703 hESC-enriched enhancers (D). In the figure (D) the average numbers of hESC-enriched enhancers are shown for sub-groups of TADs harboring 0; 1; 2; and at least 5 HARs. P values designate the statistical significance of the differences of the hESC enhancer numbers between the neighboring sub-groups of TADs. Calculations of the HAR numbers located within TADs were performed after converting the genomic coordinates of 3,127 TADs (Dixon et al., 2012) from the hg18 to hg19 release of the human reference genome database using the LiftOver algorithm.

**Figure 2.** Markedly distinct structural features of SEs and TADs in genomes of human and mouse ESCs.

A. Size distribution analyses revealed significantly increased size of SEs and decreased size of TADs in the hESC genome compared with mouse. Top two panels show the median sizes of SEs (top left panel) and TADs (top right panel) sorted in a descending order based on individual TAD sizes and segregated into ten sub-groups at 10% increments. Bottom two panels show the median (bottom left panel) and average (bottom right panel) sizes of TADs for individual chromosomes of human and mouse ESC genomes. Note that median sizes of TADs on 17 of 20 (85%) chromosomes in the hESC genome (except chr7, chr13, and chr14) are smaller compared to median TAD sizes on chromosomes in the mESC genome.

B. Numbers of both SEs (top left panel) and TADs (top right panel) are increased in the hESC genome compared with mouse. Bottom two panels illustrate that numbers of TADs are decreased in differentiated cells compared to ESC both in humans (bottom left panel) and mice (bottom right panel).

**Figure 3.** Placements of hESC enhancers and HARs within TADs are directly correlated with the size of TADs in the hESC genome.

A. Significant direct correlations between the size of TADs and SE's span defined as a number of bp between the two most distant SEs located within a given TAD in the hESC genome (top two panels). Top left panel shows a correlation profile between the size of 504 TADs and genomic span of 642 SEs located within TADs. Top right panel shows a correlation profile between the size of 103 TADs harboring at least two SEs and genomic span of 241 SEs residing within TADs. Percentiles within shaded areas indicate the percent of TADs containing HSGRL within a designated set, which increases concomitantly with the increasing quantity of SEs located within TADs and larger TAD size. Note that similar trends were observed in the mESC genome, however, the correlation coefficient values were not statistically significant (bottom two panels).

B. Patterns of significant direct correlations between the size of 147 revTADs and numbers of hESC-enriched enhancers (top left panel) and numbers of HARs (top right panel) located within the revTADs. The bottom left panel illustrates the profile of significant correlation between the numbers of hESC-enriched enhancers and HARs residing within the revTADs (direct correlation; bottom left panel). In contrast, the bottom right panel shows the previously reported inverse correlation between the numbers of HSTFBS and HARs residing within the revTADs (Glinsky, 2016).

C. Genome-wide, there is a highly significant direct correlation between the numbers of hESC-enriched enhancers located within TADs and the average size of corresponding TADs harboring hESC-enriched enhancers. The numbers of hESC-enriched enhancers located within each individual TAD were quantified and TADs harboring 0 to 9 enhancers were segregated into subgroups harboring the same numbers of enhancers. The numbers of TADs in each subgroup are indicated. Eighty-one TADs harboring ten or more hESC-enriched enhancers (range 10-30) were segregated into one subgroup with the average enhancers' content of 12.5 per TAD. The average TAD sizes were computed for each subgroup of TADs and the results were plotted to assess the correlation pattern.

**Figure 4.** Mechanisms of HSGRL-mediated effects on principal regulatory structures of the interphase chromatin contributing to the evolution of genomic regulatory networks.

Collectively, the ensemble of these structural changes facilitated by the targeted placements and retention of distinct families of HSGRL at specific genomic locations would enable the emergence of new SED structures and remodeling of existing TADs to increase the complexity and enhance the precision of genomic regulatory networks. See text for details and relevant references.

**Figure 5.** Clusters of Alu elements in the vicinity of putative DNA bending sites near the borders of SEDs and TADs.

A. Clusters of Alu elements near the borders of the ID3 SED on chr1:23,878,033-23,894,300 (16,268 bp). Dotted lines depict the genomic positions of the overlapping CTCF/cohesin-binding sites, interactions of which form the anchor base of the ID3 SED.

B. Clusters of Alu elements near the NANOG SED left border on chr12:7,864,594-7,869,500 (4,907 bp).

C. Clusters of Alu elements near the NANOG SED right border on chr12:8,012,400-8,017,400 (5,001 bp).

Arrows in the figures (B) and )C) point the overlapping CTCF/cohesin-binding sites, interactions between which form the anchor base of the NANOG SED.

Note that clusters of closely-spaced sequences of at least three Alu elements belonging to most ancient AluJ (~65 million years old), second oldest AluS (~30 million years old), and currently active modern AluY sub-families are observed, suggesting that placement and/or retention of Alu elements at these sites were occurring for millions of years and continues at the present time (see text for details).

**Figure 6.** Conservation patterns of HSGRL in individual human genomes.

Conservation of HSGRL in individual genomes of 3 Neanderthals, 12 Modern Humans, and the 41,000-year old Denisovan genome was carried-out by direct comparisons of corresponding sequences retrieved from individual genomes and the human genome reference database (http://genome.ucsc.edu/Neandertal/). Full-length sequence alignments with no gaps of the individual human genome sequences to the corresponding sequences in the human genome reference databases were accepted as the evidence of sequence conservation in the individual human genome.

A. Top two panels show conservation patterns of HSGRL located within the revTADs for 57 HARs and associated 75 CTCF-binding sites (top left panel); 22 HARs harboring 23 LMNB1-binding sites and 23 LMNB1-binding sites residing within 22 HARs (top right panel). Bottom two panels show conservation patterns of 90 HARs harboring 93 LMB1-binding sites (bottom left panel) and 55 HARs harboring 55 CTCF-binding sites and 55 CTCF-binding sites located within 55 HARs (bottom right panel).

B. Top two panels show conservation patterns of 123 HARs associated with 127 high-confidence overlapping CTCF/RAD21-binding sites (top left pane) and 127 high-confidence overlapping CTCF/RAD21-binding sites associated with 123 HARs (top right panel). Bottom two panels show conservation patterns of 69 CTCF/RAD21-binding sites conserved in all Neandertals' genome (bottom left panel) and 152 human-specific NANOG-binding sites highly-conserved in individual human genomes (bottom right panel). Note that conservation of all HSGRL are consistently at the lowest level in the Neanderthals' genome, suggesting that creation and/or retention rates of HSGRL are enhanced in Modern Humans. However, HSGRL sequences manifesting the relatively high conservation levels in the Neanderthals' genome appear most conserved in the individual genomes of Modern Humans as well, including the 41,000-year old Denisovan genome (bottom left panel). Sequences of HSGRL with assigned biochemical functions, e.g., specific TFBS or Lamin B1 (LMNB1)-binding sites, which are residing within HARs exhibit markedly higher conservation levels compared to sequences of HARs harboring the corresponding HSGRL. This conclusion remains valid for HSGRL sequences with assigned specific biochemical or biological functions that were associated with HARs by proximity placement analyses (figures A, B).

C. Direct correlations of conservation profiles of DNA sequences of distinct HSGRL in individual human genomes (top two panels) and markedly different values of genomic fitness scores (GFS) integrating into a single numerical value sequence conservation data of 909 HSGRL in individual human genomes (bottom panel). Note that values of GFS are inversely correlated with the variation coefficients of GFS values in individual human genomes, consistent with the hypothesis that GFS reflects the intrinsic property of an individual genome to create and/or retain the HSGRL.

**Table 1.** Association of hESC super-enhancers with human-specific regulatory loci

| Genomic Features* | TADs | H1 hESC-SE | hsTFBS | HARs |
|---|---|---|---|---|
| Genome | 3,062 | 642 | 3803 | 2745 |
| HARs + SE | 168 | 212 | 0 | 406 |
| Enrichment Factor | 1.0 | 6.0 | NA | 2.8 |
| HSTFBS + SE | 56 | 72 | 146 | 0 |
| Enrichment Factor | 1.0 | 6.1 | 2.1 | NA |
| HARs + HSTFBS + SE | 55 | 85 | 154 | 158 |
| Enrichment Factor | 1.0 | 7.4 | 2.3 | 3.3 |
| SE + HSGRL | 279 | 369 | 300 | 564 |

HARs, Human Accelerated Regions; HSTFBS, human-specific Transcription Factor-Binding Sites; SE, Super Enhancers; HSGRL, human-specific genomic regulatory loci; H1 hESC, H1 human Embryonic Stem Cells; NA, not applicable; * - genomic coordinates of the hg19 human genome reference database;

**Table 2.** Size distribution analysis of super-enhancers in human and mouse ESC

| Category | hESC super-enhancers | | mESC super-enhancers | | Fold enrichment in hESC |
|---|---|---|---|---|---|
| Size, bp | Number | Percent | Number | Percent | |
| > 30,000 | 29 | 4.2 | 4 | 1.7 | 7.25 |
| >20,000 | 102 | 14.9 | 28 | 12.1 | 3.64 |
| >10,000 | 331 | 48.4 | 108 | 46.8 | 3.06 |
| >5,000 | 554 | 81.0 | 154 | 66.7 | 3.60 |
| >2,000 | 683 | 99.9 | 193 | 83.5 | 3.54 |
| 1,000-2,000 | 0 | 0.0 | 21 | 9.1 | 0.00 |
| <1,000 | 1 | 0.1 | 17 | 7.4 | 0.06 |
| Total | 684 | 100.0 | 231 | 100.0 | 2.96 |

hESC, human embryonic stem cells; mESC, mouse embryonic stem cells

**Table 3.** Size distribution analysis of topologically-associating domains in genomes of human and mouse ESC

| Category | hESC TADs | | mESC TADs | | Enrichment/Depletion in hESC |
|---|---|---|---|---|---|
| Size, Kbp | Number | Percent | Number | Percent | |
| > 3,000 | 23 | 0.7 | 55 | 2.5 | 0.42 |
| >2,000 | 160 | 5.1 | 255 | 11.6 | 0.63 |
| >1,000 | 955 | 30.5 | 982 | 44.6 | 0.97 |
| >500 | 2154 | 68.9 | 1745 | 79.3 | 1.23 |
| >200 | 3054 | 97.7 | 2166 | 98.5 | 1.41 |
| >100 | 3125 | 99.9 | 2200 | 100.0 | 1.42 |
| <100 | 2 | 0.1 | 0 | 0.0 | NA |
| Total | 3127 | 100.0 | 2200 | 100.0 | 1.42 |

TADs, topologically-associating domains

**Table 4.** Estimates of creation time periods during evolution of SEs and enhancers in the hESC genome

**1.1.** Estimates of super-enhancers (SE) creation time in the human ESC genome

| Number of SE in hESC genome | Shared with mESC | All new SE in hESC* | New SE on conserved sequences* | Primate-specific SE sequences* | Human-specific SE sequences** |
|---|---|---|---|---|---|
| 684 | 25 (3.7%) | 659 | 573 | 83 | 3 |
| Number of new super-enhancers per 100,000 years | | 0.88 | 0.76 | 0.11 | 0.81 |
| Number of years for creation of one super-enhancer | | 113,636 | 131,579 | 909,091 | 123,457 |

**1.2.** Estimates of enhancers' creation time in the human ESC genome

| Number of enhancers in hESC genome | Shared with mESC | All new enhancers* | New enhancers on conserved sequences* | Primate-specific enhancers sequences* | Human-specific enhancers sequences** |
|---|---|---|---|---|---|
| 7006 | 682 (9.7%) | 6,324 | 4,714 | 1,420 | 190 |
| Number of new enhancers per 100,000 years | | 8.4 | 6.3 | 1.9 | 51 |
| Number of years for creation of one enhancer | | 11,905 | 15,873 | 52,632 | 1,961 |

* Calculated based on estimates of Humans & Chimpanzees split 13 million years ago and 88 million years from Euarchonta & Glires (Gliriformes) split resulting in the estimated evolutionary timeline of ~75 million years;
** Calculated based on estimates of Modern Humans & Neanderthals split 370,000 years ago;

Estimates for human-specific SE and enhancers were calculated based on the assumption that all human-specific sequences emerged after Modern Humans & Neanderthals split 370,000 years ago;

Estimates of creation time of primate-specific SE and enhancers were based on the assumption that all primate-specific sequences emerged before Modern Humans/Chimpanzees split 13 million years ago.

**Figure 1.**

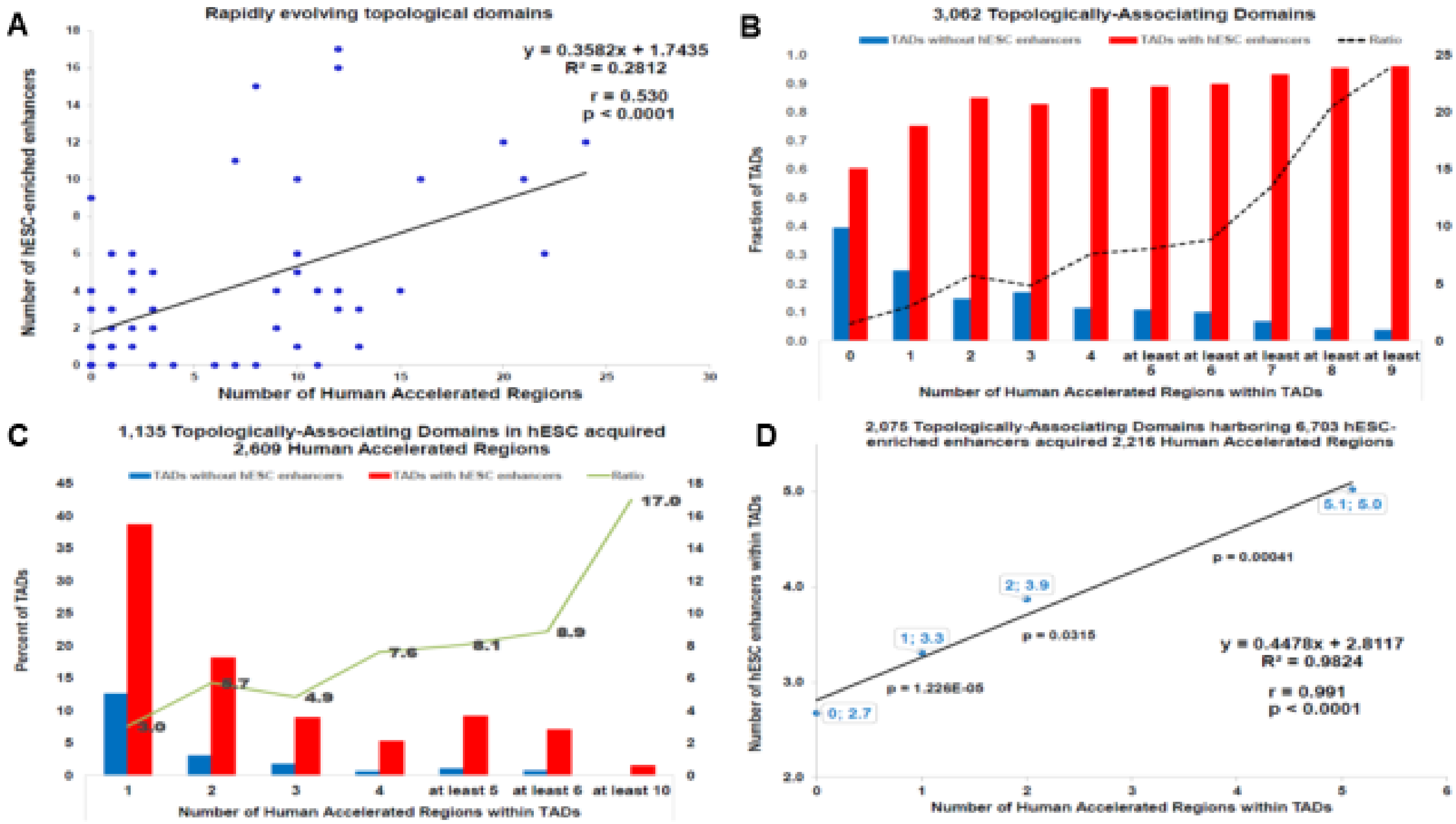
A
Rapidly evolving topological domains
y = 0.3582x + 1.7435
R² = 0.2812
r = 0.530
p < 0.0001
Number of hESC-enriched enhancers
Number of Human Accelerated Regions
B
3,062 Topologically-Associating Domains
TADs without hESC enhancers
TADs with hESC enhancers
Ratio
Fraction of TADs
0 1 2 3 4 at least 5 at least 6 at least 7 at least 8 at least 9
Number of Human Accelerated Regions within TADs
C
1,135 Topologically-Associating Domains in hESC acquired 2,609 Human Accelerated Regions
TADs without hESC enhancers
TADs with hESC enhancers
Ratio
Percent of TADs
3.0
5.7
4.9
7.6
8.1
8.9
17.0
1 2 3 4 at least 5 at least 6 at least 10
Number of Human Accelerated Regions within TADs
D
2,075 Topologically-Associating Domains harboring 6,703 hESC-enriched enhancers acquired 2,216 Human Accelerated Regions
0; 2.7
1; 3.3
2; 3.9
5.1; 5.0
p = 1.226E-05
p = 0.0315
p = 0.00041
y = 0.4478x + 2.8117
R² = 0.9824
r = 0.991
p < 0.0001
Number of hESC enhancers within TADs
Number of Human Accelerated Regions within TADs

**Figure 2.**

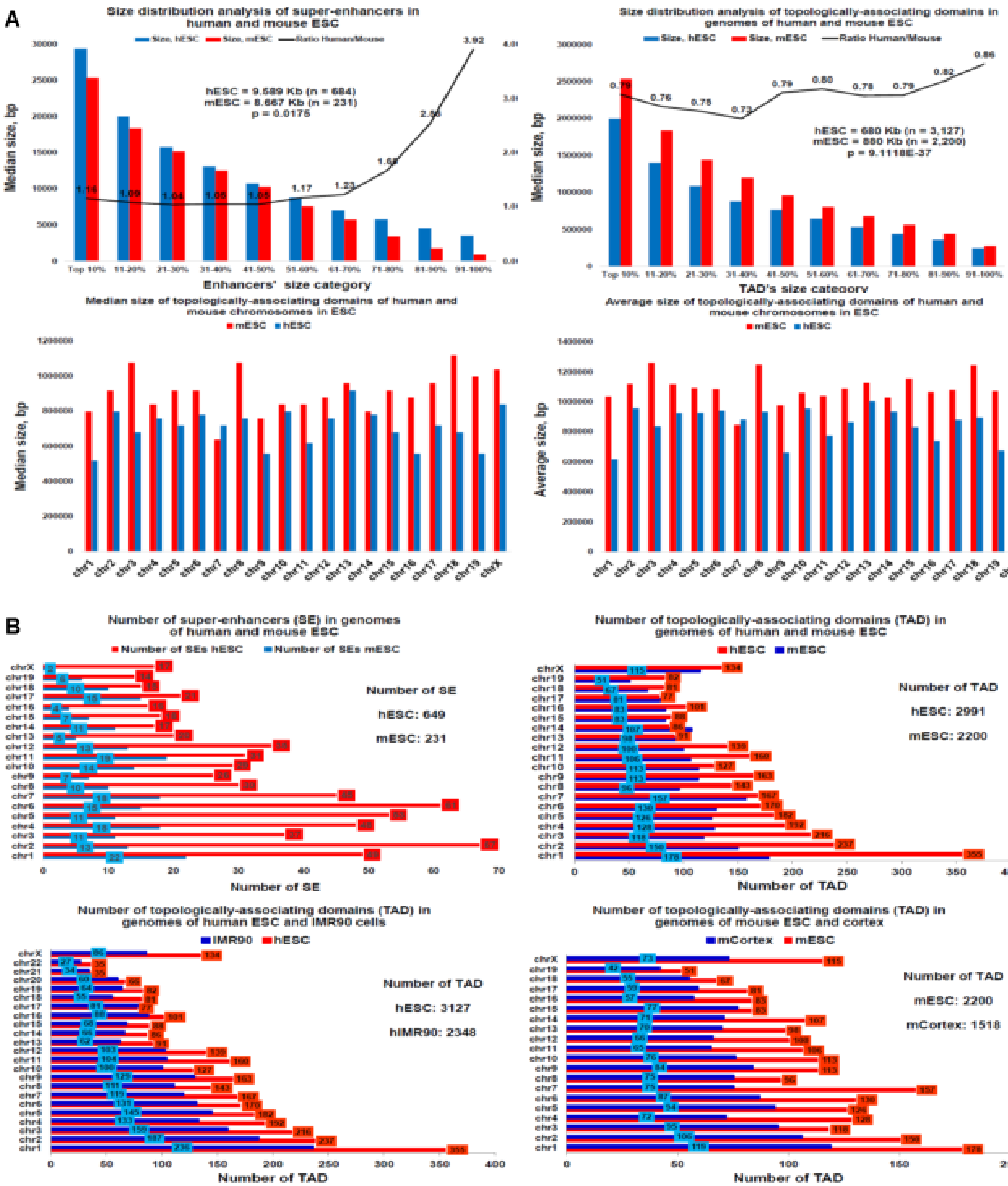
A
Size distribution analysis of super-enhancers in human and mouse ESC
Size, hESC
Size, mESC
Ratio Human/Mouse
hESC = 9.589 Kb (n = 684)
mESC = 8.667 Kb (n = 231)
p = 0.0175
Median size, bp
Enhancers' size category
Size distribution analysis of topologically-associating domains in genomes of human and mouse ESC
hESC = 680 Kb (n = 3,127)
mESC = 880 Kb (n = 2,200)
p = 9.1118E-37
TAD's size category
Median size of topologically-associating domains of human and mouse chromosomes in ESC
mESC
hESC
Average size of topologically-associating domains of human and mouse chromosomes in ESC
Average size, bp
B
Number of super-enhancers (SE) in genomes of human and mouse ESC
Number of SEs hESC
Number of SEs mESC
Number of SE
hESC: 649
mESC: 231
Number of topologically-associating domains (TAD) in genomes of human and mouse ESC
Number of TAD
hESC: 2991
mESC: 2200
Number of topologically-associating domains (TAD) in genomes of human ESC and IMR90 cells
IMR90
hESC: 3127
hIMR90: 2348
Number of topologically-associating domains (TAD) in genomes of mouse ESC and cortex
mCortex
mESC: 2200
mCortex: 1518

**Figure 3.**

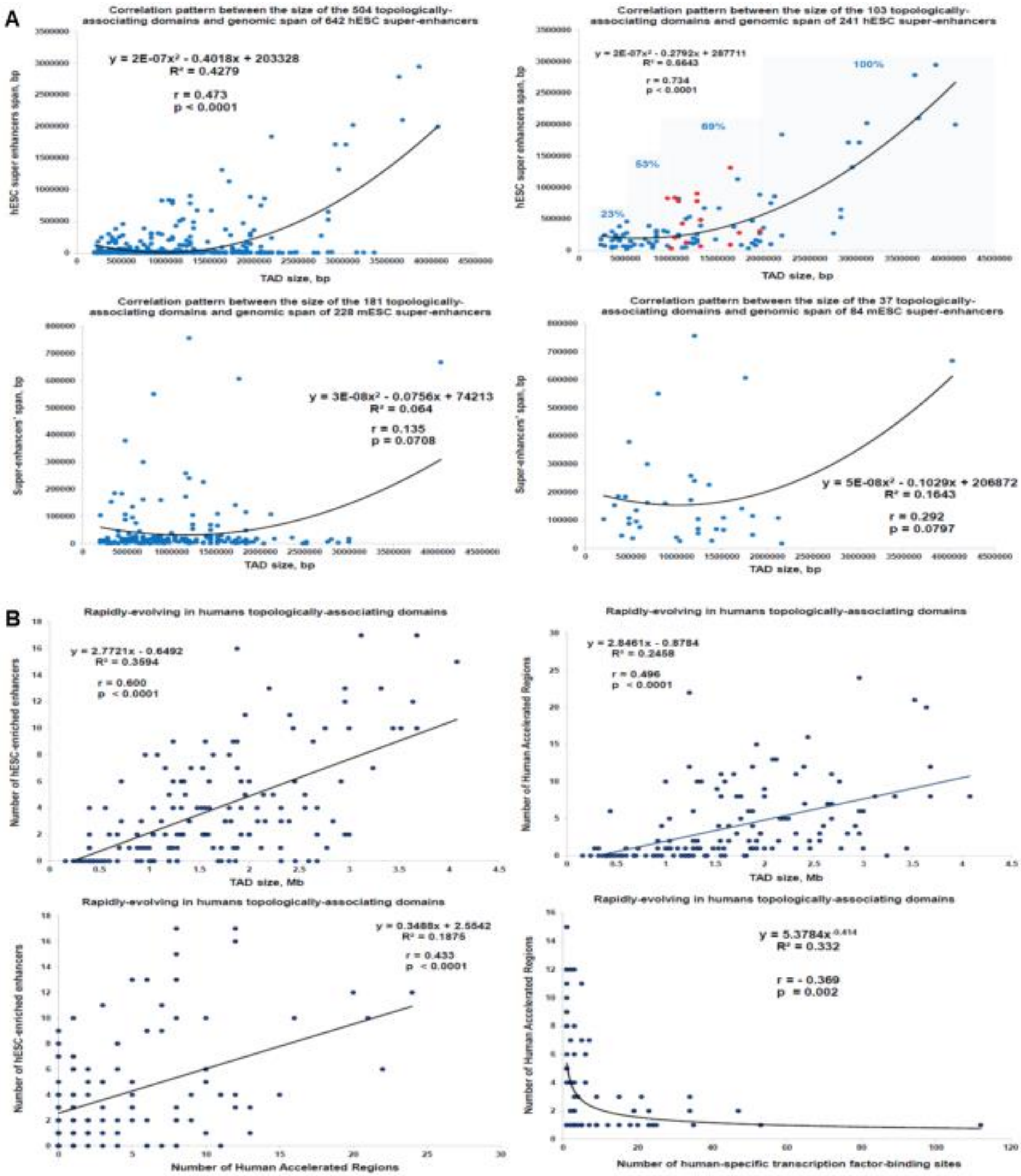
A
Correlation pattern between the size of the 504 topologically-associating domains and genomic span of 642 hESC super-enhancers
y = 2E-07x² - 0.4018x + 203328
R² = 0.4279
r = 0.473
p < 0.0001
hESC super enhancers span, bp
TAD size, bp
Correlation pattern between the size of the 103 topologically-associating domains and genomic span of 241 hESC super-enhancers
y = 2E-07x² - 0.2792x + 287711
R² = 0.6643
r = 0.734
p < 0.0001
100%
69%
53%
23%
hESC super enhancers span, bp
TAD size, bp
Correlation pattern between the size of the 181 topologically-associating domains and genomic span of 228 mESC super-enhancers
y = 3E-08x² - 0.0756x + 74213
R² = 0.064
r = 0.135
p = 0.0708
Super-enhancers' span, bp
TAD size, bp
Correlation pattern between the size of the 37 topologically-associating domains and genomic span of 84 mESC super-enhancers
y = 5E-08x² - 0.1029x + 206872
R² = 0.1643
r = 0.292
p = 0.0797
Super-enhancers' span, bp
TAD size, bp
B
Rapidly-evolving in humans topologically-associating domains
y = 2.7721x - 0.6492
R² = 0.3594
r = 0.600
p < 0.0001
Number of hESC-enriched enhancers
TAD size, Mb
Rapidly-evolving in humans topologically-associating domains
y = 2.8461x - 0.8784
R² = 0.2458
r = 0.496
p < 0.0001
Number of Human Accelerated Regions
TAD size, Mb
Rapidly-evolving in humans topologically-associating domains
y = 0.3488x + 2.5642
R² = 0.1875
r = 0.433
p < 0.0001
Number of hESC-enriched enhancers
Number of Human Accelerated Regions
Rapidly-evolving in humans topologically-associating domains
y = 5.3784x^-0.414
R² = 0.332
r = - 0.369
p = 0.002
Number of Human Accelerated Regions
Number of human-specific transcription factor-binding sites

C

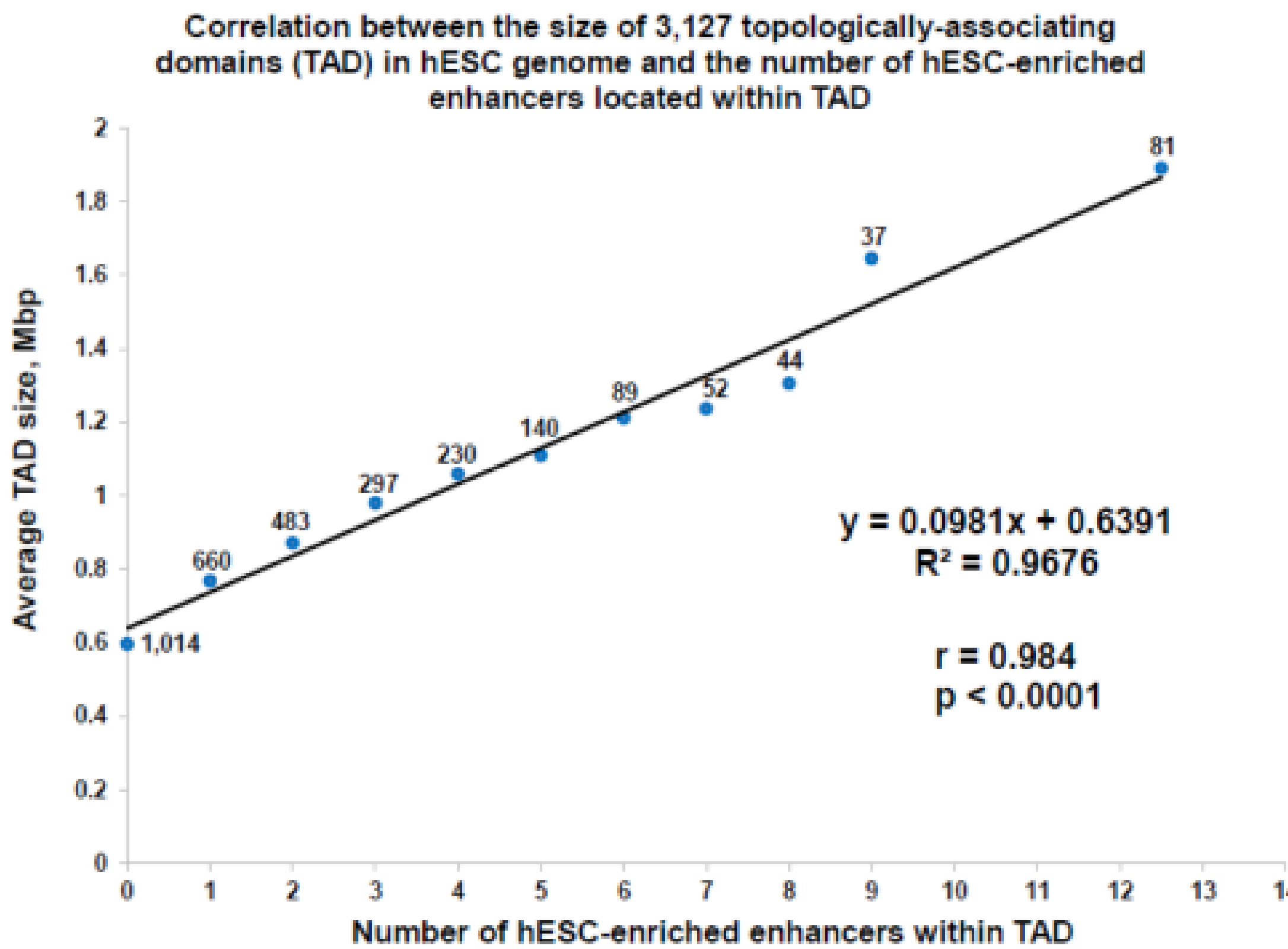

Correlation between the size of 3,127 topologically-associating domains (TAD) in hESC genome and the number of hESC-enriched enhancers located within TAD
Average TAD size, Mbp
Number of hESC-enriched enhancers within TAD
y = 0.0981x + 0.6391
R² = 0.9676
r = 0.984
p < 0.0001
1,014
660
483
297
230
140
89
52
44
37
81

**Figure 4.**

### Mechanisms of HSGRL-mediated effects on principal regulatory structures of the interphase chromatin

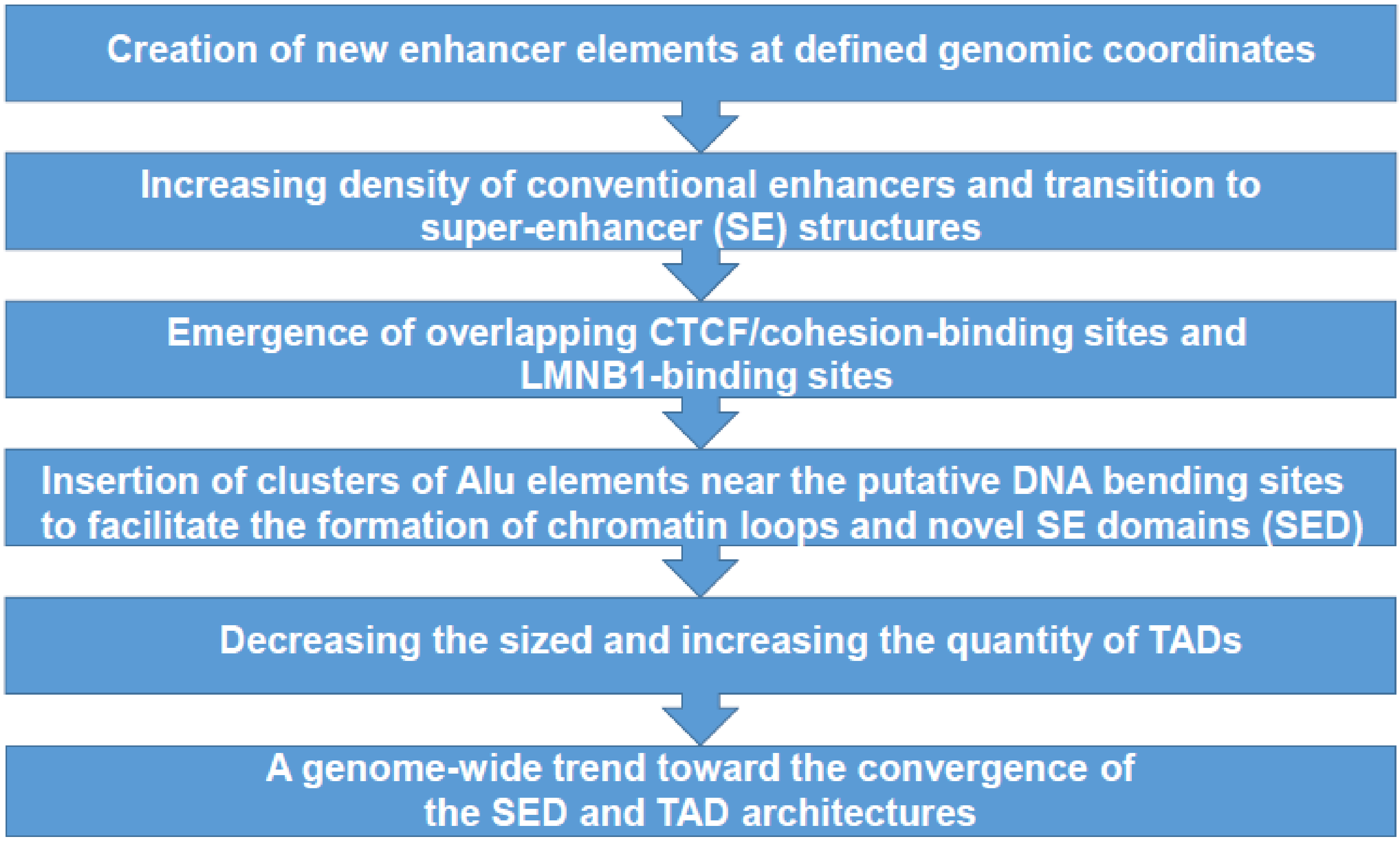

**Figure 5.**

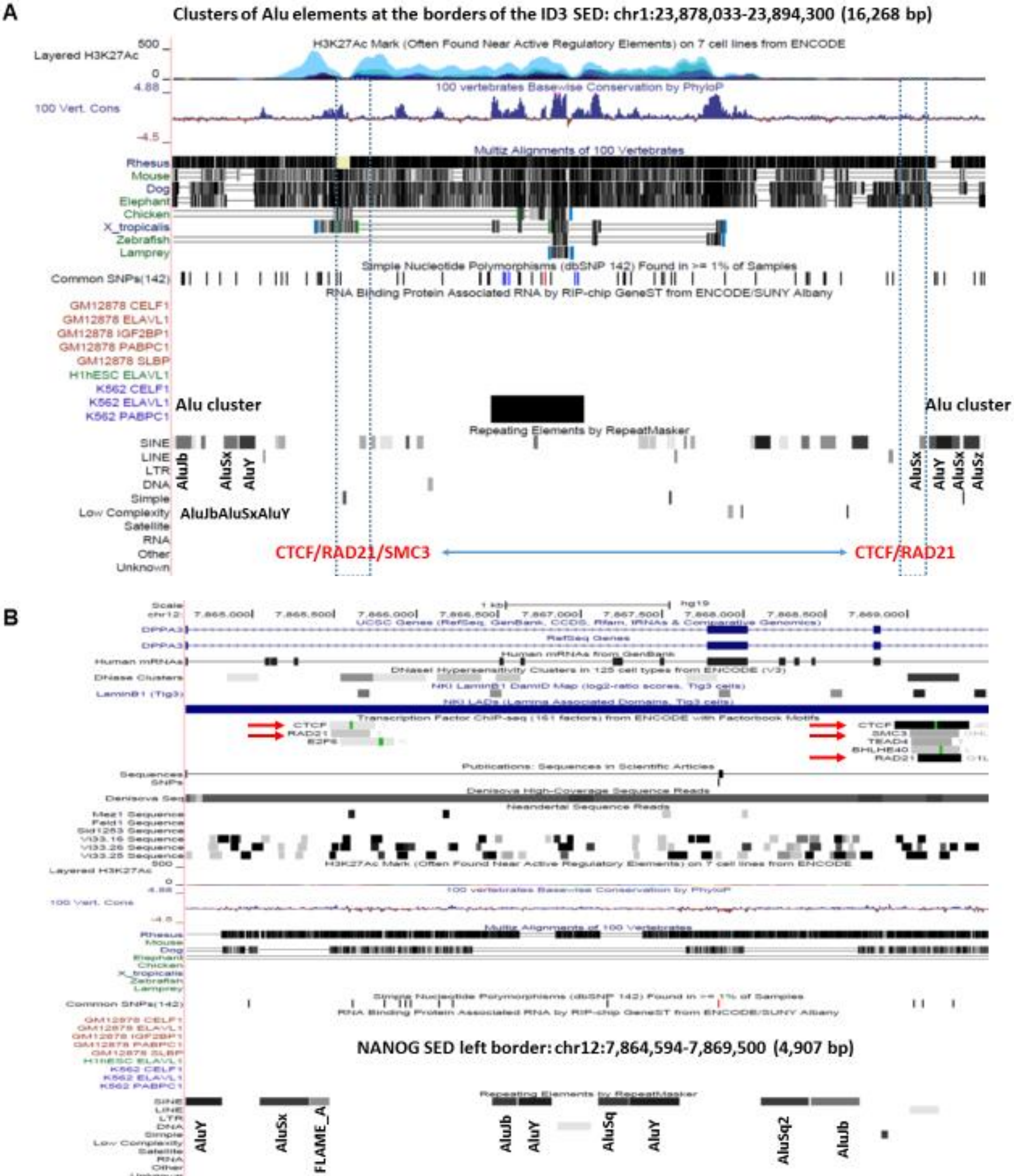

A
Clusters of Alu elements at the borders of the ID3 SED: chr1:23,878,033-23,894,300 (16,268 bp)
Layered H3K27Ac
500
0
H3K27Ac Mark (Often Found Near Active Regulatory Elements) on 7 cell lines from ENCODE
4.88
100 Vert. Cons
-4.5
100 vertebrates Basewise Conservation by PhyloP
Multiz Alignments of 100 Vertebrates
Rhesus
Mouse
Dog
Elephant
Chicken
X_tropicalis
Zebrafish
Lamprey
Simple Nucleotide Polymorphisms (dbSNP 142) Found in >= 1% of Samples
Common SNPs(142)
RNA Binding Protein Associated RNA by RIP-chip GeneST from ENCODE/SUNY Albany
GM12878 CELF1
GM12878 ELAVL1
GM12878 IGF2BP1
GM12878 PABPC1
GM12878 SLBP
H1hESC ELAVL1
K562 CELF1
K562 ELAVL1
K562 PABPC1
Alu cluster
Alu cluster
Repeating Elements by RepeatMasker
SINE
LINE
LTR
DNA
Simple
Low Complexity
Satellite
RNA
Other
Unknown
AluJb
AluSx
AluY
AluJbAluSxAluY
AluSx
AluY
AluSx
AluSz
CTCF/RAD21/SMC3
CTCF/RAD21
B
Scale
chr12:
1 kb
hg19
7,865,000
7,865,500
7,866,000
7,866,500
7,867,000
7,867,500
7,868,000
7,868,500
7,869,000
DPPA3
RefSeq Genes
Human mRNAs
DNase Clusters
LaminB1 (Tig3)
CTCF
RAD21
E2F6
CTCF
SMC3
TEAD4
BHLHE40
RAD21
Sequences
SNPs
Denisova Seq
Layered H3K27Ac
100 Vert. Cons
Rhesus
Mouse
Dog
Common SNPs(142)
NANOG SED left border: chr12:7,864,594-7,869,500 (4,907 bp)
SINE
LINE
LTR
DNA
Simple
Low Complexity
Satellite
RNA
Other
Unknown
AluY
AluSx
FLAME_A
AluJb
AluY
AluSq
AluY
AluSq2
AluJb

C

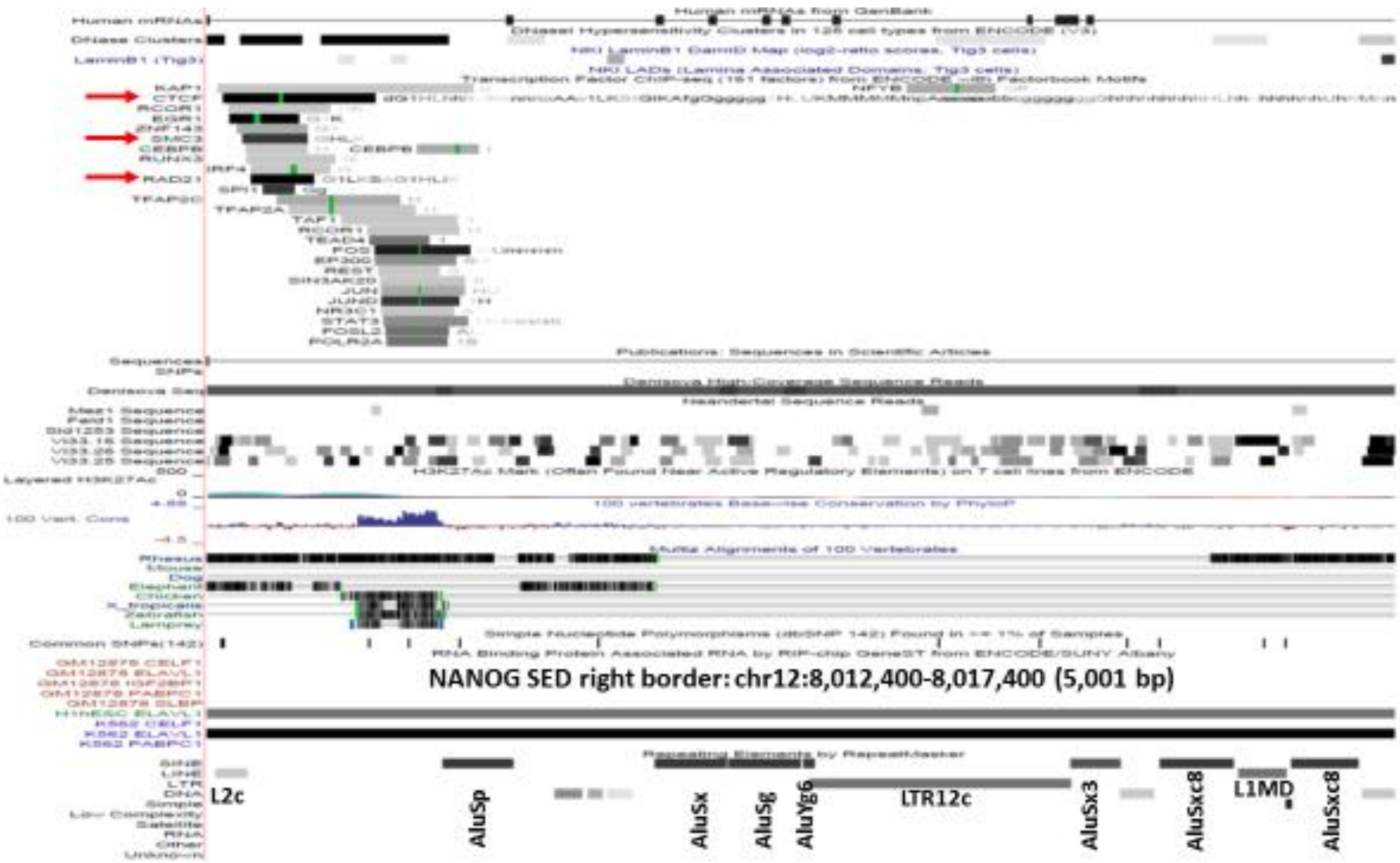

NANOG SED right border: chr12:8,012,400-8,017,400 (5,001 bp)
L2c
AluSp
AluSx
AluSg
AluYg6
LTR12c
AluSx3
AluSxc8
L1MD
AluSxc8

**Figure 6.**

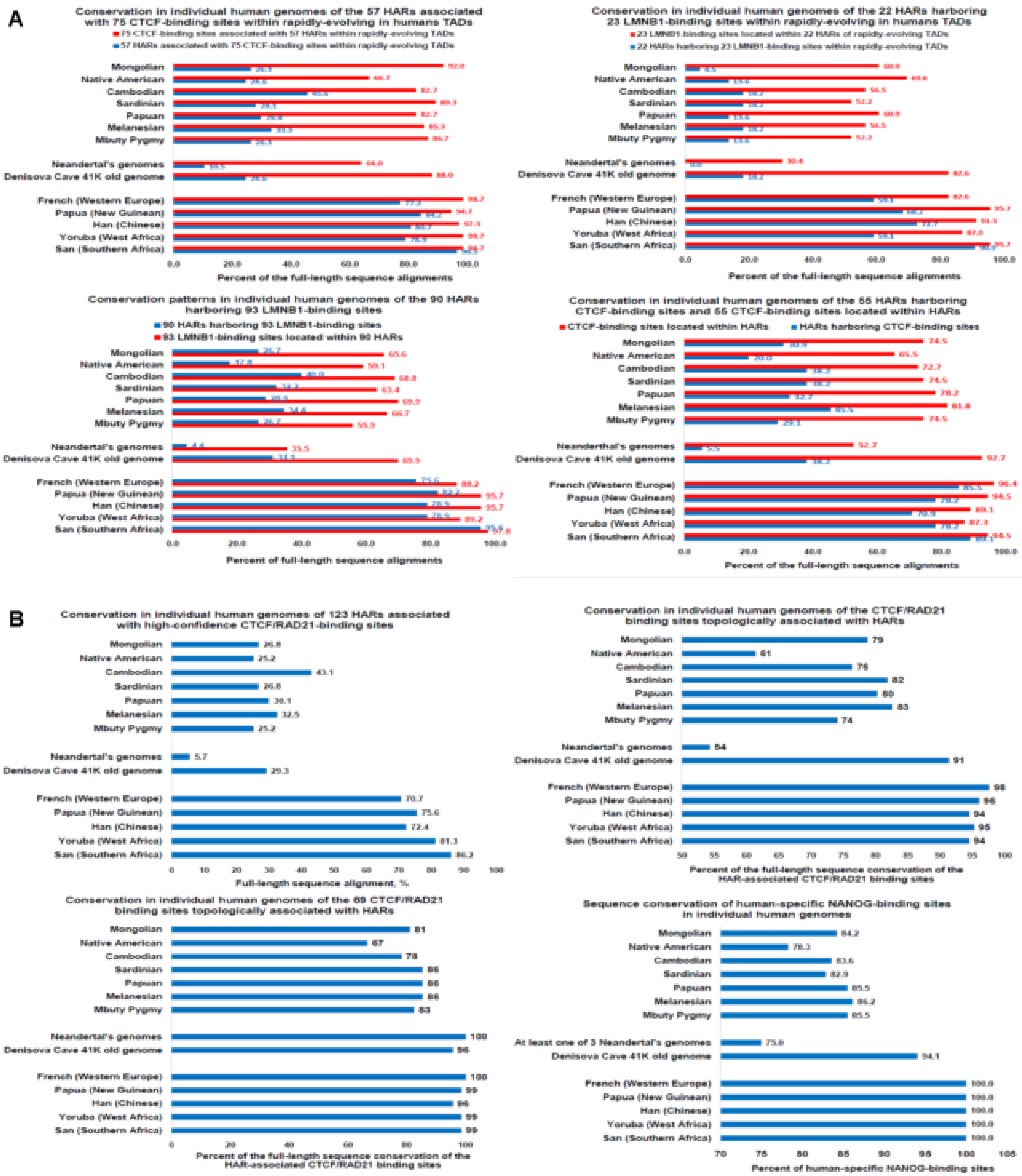
A
Conservation in individual human genomes of the 57 HARs associated with 75 CTCF-binding sites within rapidly-evolving in humans TADs
Conservation in individual human genomes of the 22 HARs harboring 23 LMNB1-binding sites within rapidly-evolving in humans TADs
Conservation patterns in individual human genomes of the 90 HARs harboring 93 LMNB1-binding sites
90 HARs harboring 93 LMNB1-binding sites
93 LMNB1-binding sites located within 90 HARs
Conservation in individual human genomes of the 55 HARs harboring CTCF-binding sites and 55 CTCF-binding sites located within HARs
CTCF-binding sites located within HARs
HARs harboring CTCF-binding sites
Mongolian
Native American
Cambodian
Sardinian
Papuan
Melanesian
Mbuty Pygmy
Neandertal's genomes
Denisova Cave 41K old genome
French (Western Europe)
Papua (New Guinean)
Han (Chinese)
Yoruba (West Africa)
San (Southern Africa)
Percent of the full-length sequence alignments
Percent of full-length sequence alignments
B
Conservation in individual human genomes of 123 HARs associated with high-confidence CTCF/RAD21-binding sites
Full-length sequence alignment, %
Conservation in individual human genomes of the CTCF/RAD21 binding sites topologically associated with HARs
Percent of the full-length sequence conservation of the HAR-associated CTCF/RAD21 binding sites
Conservation in individual human genomes of the 69 CTCF/RAD21 binding sites topologically associated with HARs
Sequence conservation of human-specific NANOG-binding sites in individual human genomes
At least one of 3 Neandertal's genomes
Percent of human-specific NANOG-binding sites

C

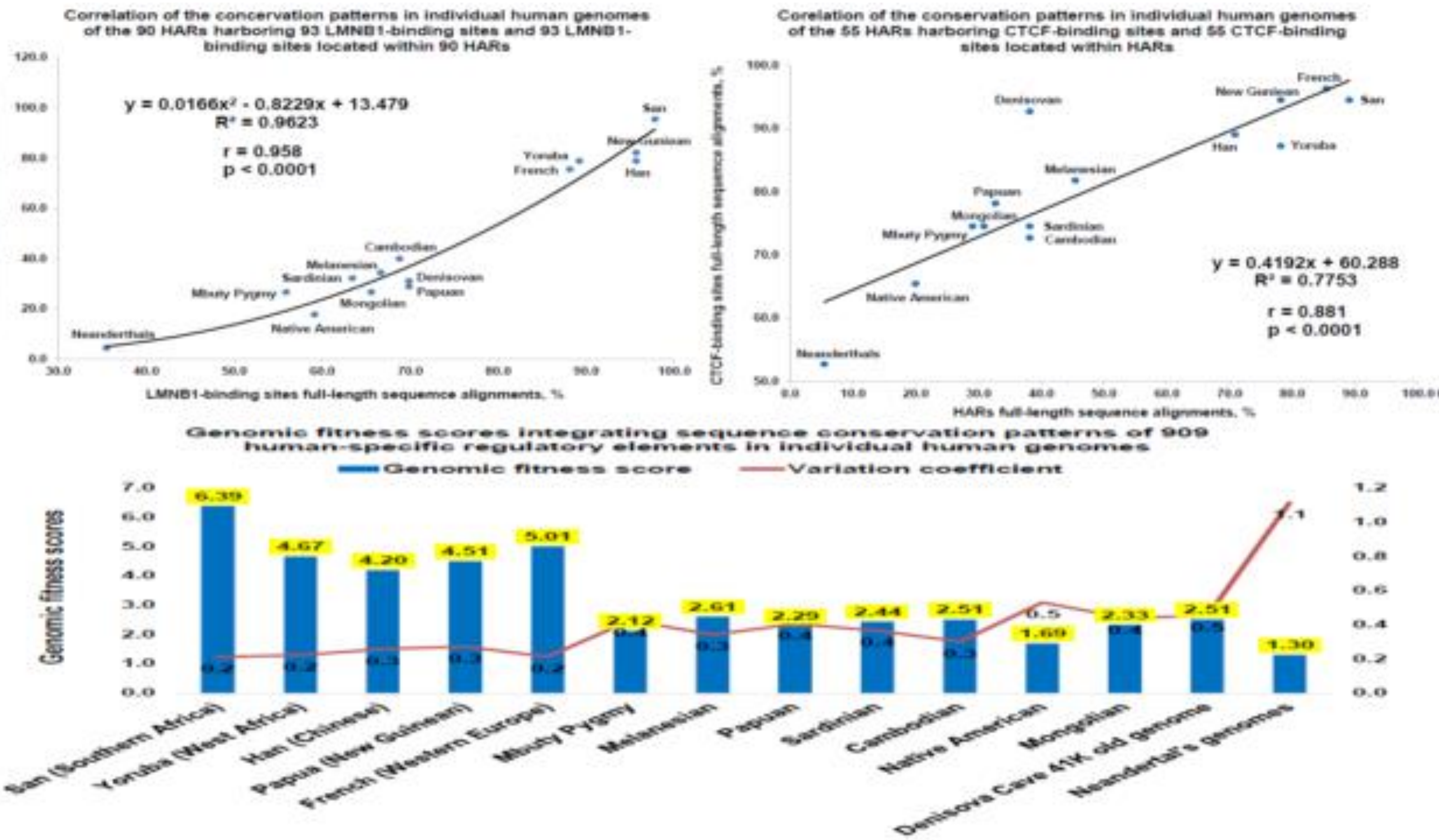

Correlation of the concervation patterns in individual human genomes of the 90 HARs harboring 93 LMNB1-binding sites and 93 LMNB1-binding sites located within 90 HARs
y = 0.0166x² - 0.8229x + 13.479
R² = 0.9623
r = 0.958
p < 0.0001
HARs full-length sequence alignments, %
LMNB1-binding sites full-length sequemce alignments, %
San
New Guinean
Yoruba
French
Han
Cambodian
Melanesian
Sardinian
Denisovan
Papuan
Mbuty Pygmy
Mongolian
Native American
Neanderthals
Corelation of the conservation patterns in individual human genomes of the 55 HARs harboring CTCF-binding sites and 55 CTCF-binding sites located within HARs
y = 0.4192x + 60.288
R² = 0.7753
r = 0.881
p < 0.0001
CTCF-binding sites full-length sequence alignments, %
HARs full-length sequence alignments, %
Genomic fitness scores integrating sequence conservation patterns of 909 human-specific regulatory elements in individual human genomes
Genomic fitness score
Variation coefficient
Genomic fitness scores
6.39
4.67
4.20
4.51
5.01
2.12
2.61
2.29
2.44
2.51
1.69
2.33
2.51
1.30
San (Southern Africa)
Yoruba (West Africa)
Han (Chinese)
Papua (New Guinean)
French (Western Europe)
Mbuty Pygmy
Melanesian
Papuan
Sardinian
Cambodian
Native American
Mongolian
Denisova Cave 41K old genome
Neanderthal's genomes

## SUPPLEMENTAL FIGURES

Supplemental Figure S1. Related to the Figure 3.
Correlation patterns between TAD size and genomic span of hESC enhancers located within the revTADs.

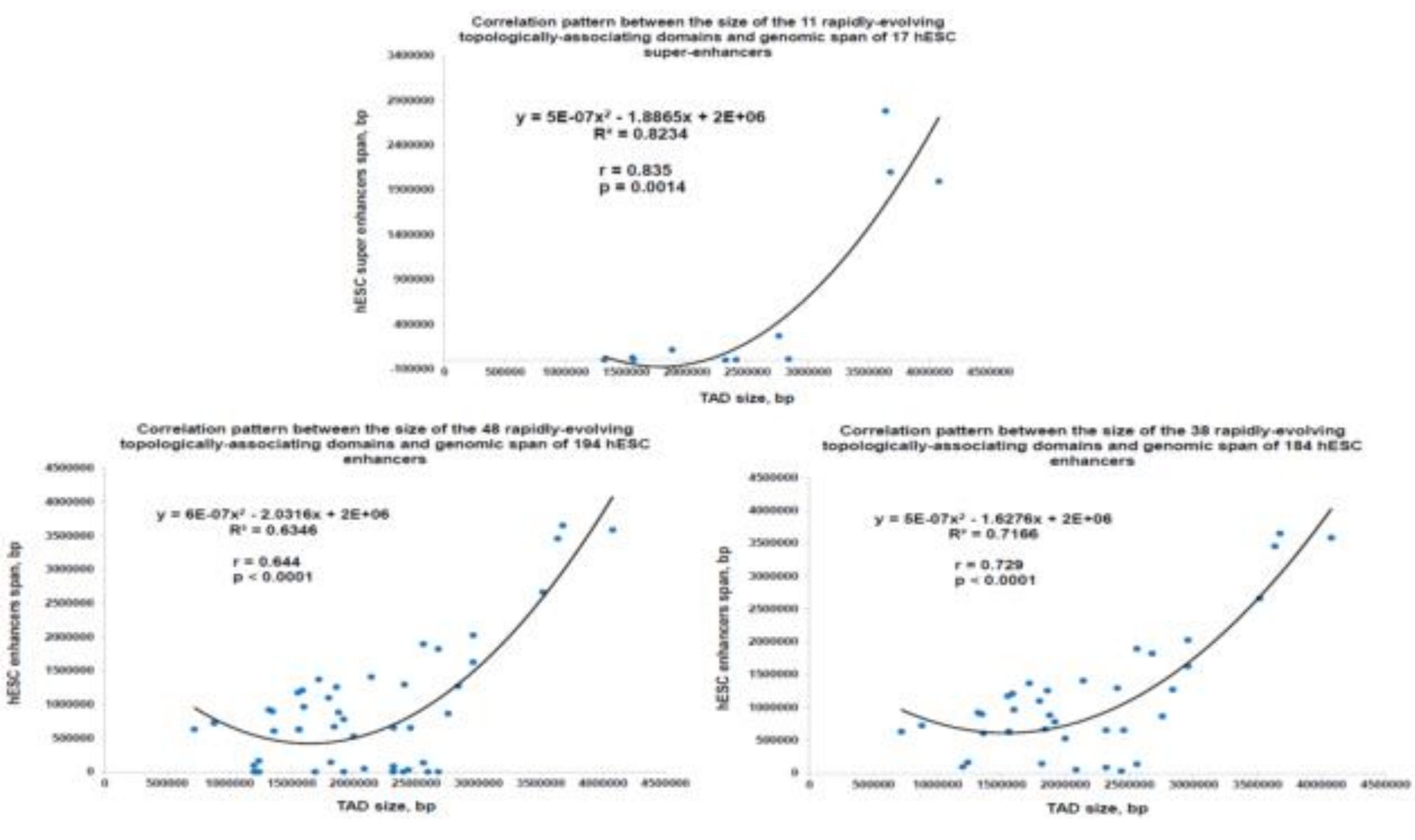

Supplemental Figure S2. Related to the Figure 5.
Expanded and zoom-in views of Alu clusters located near the putative DNA bending sites at the SED and TAD borders.

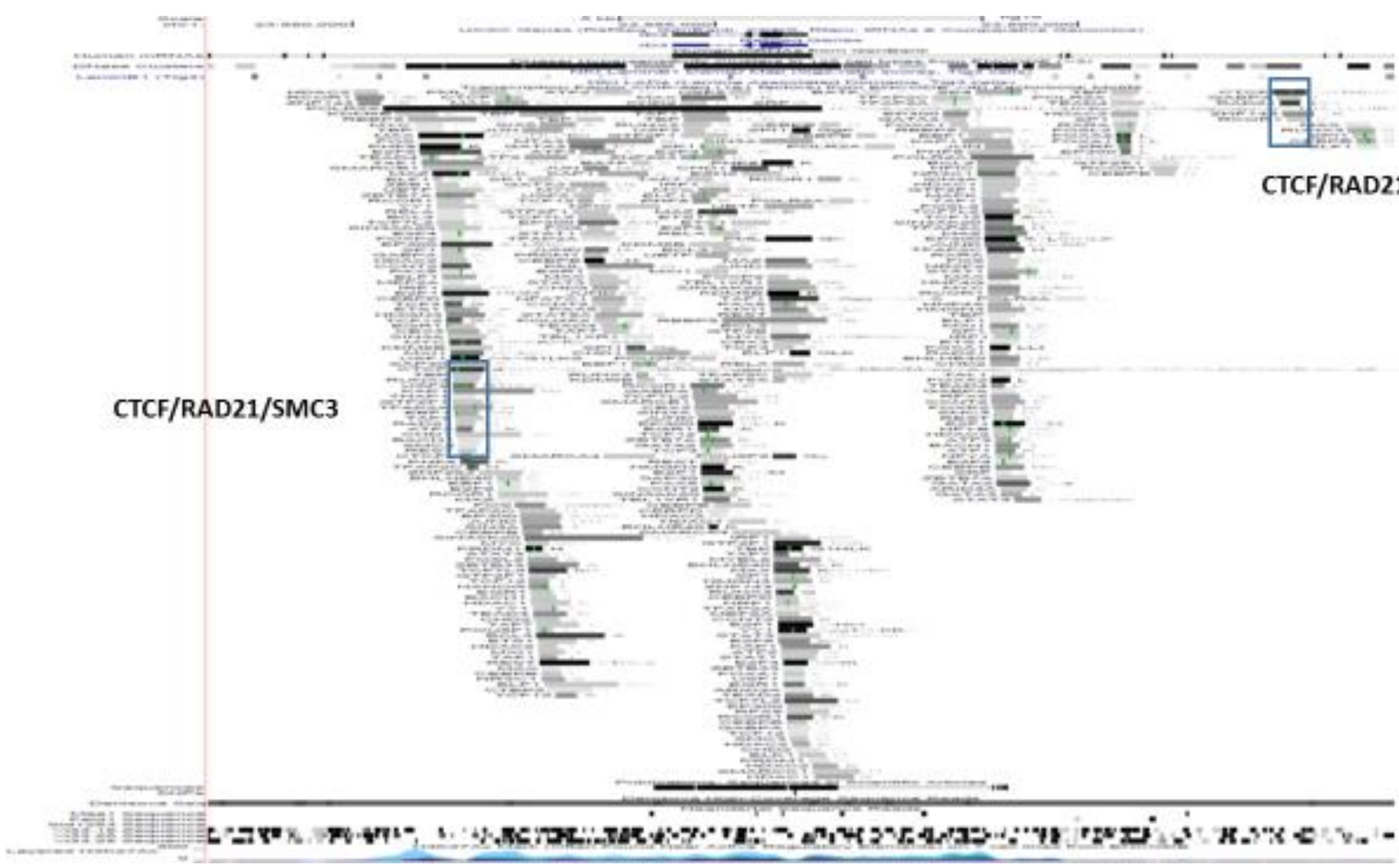

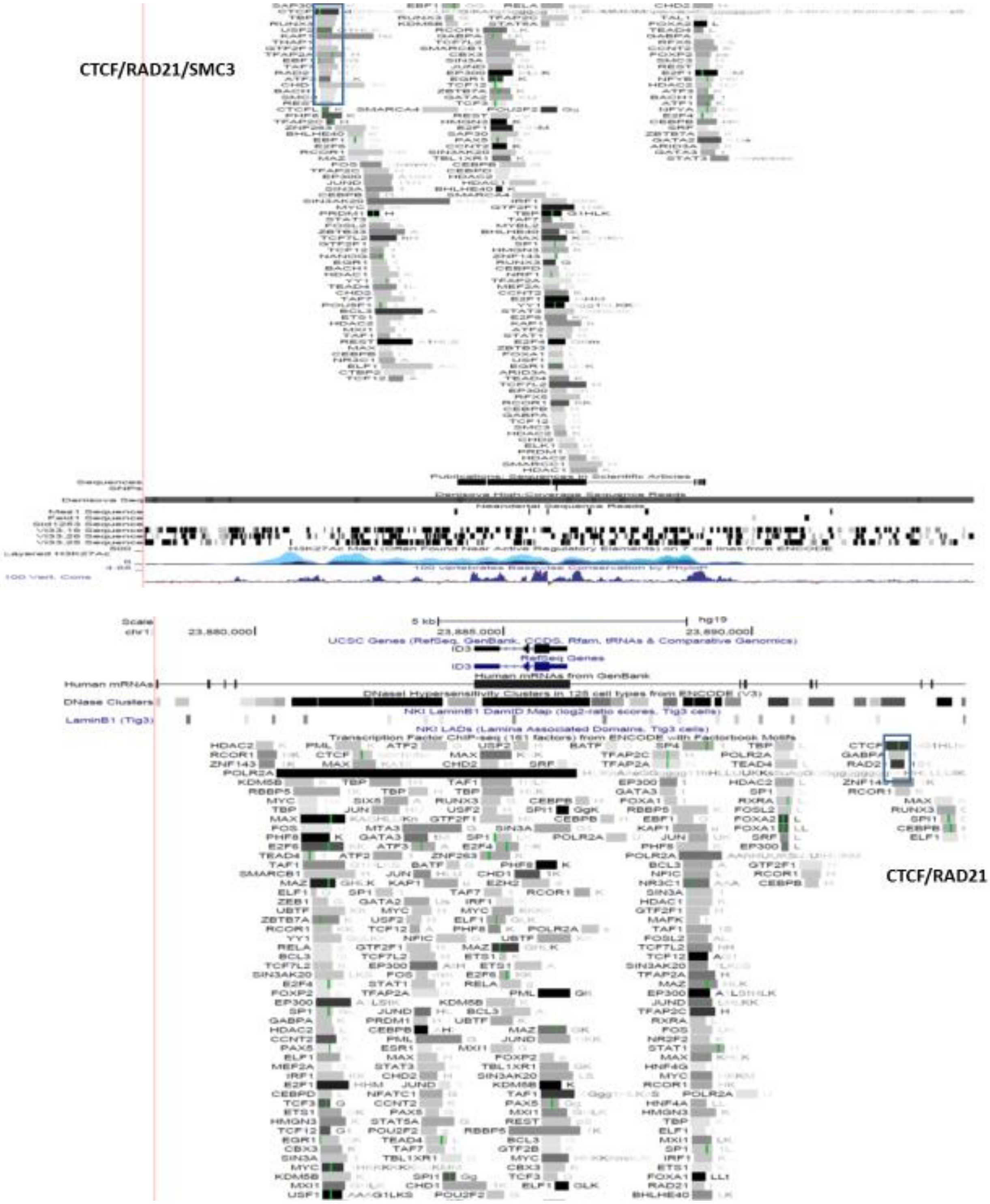
CTCF/RAD21/SMC3
CTCF/RAD21

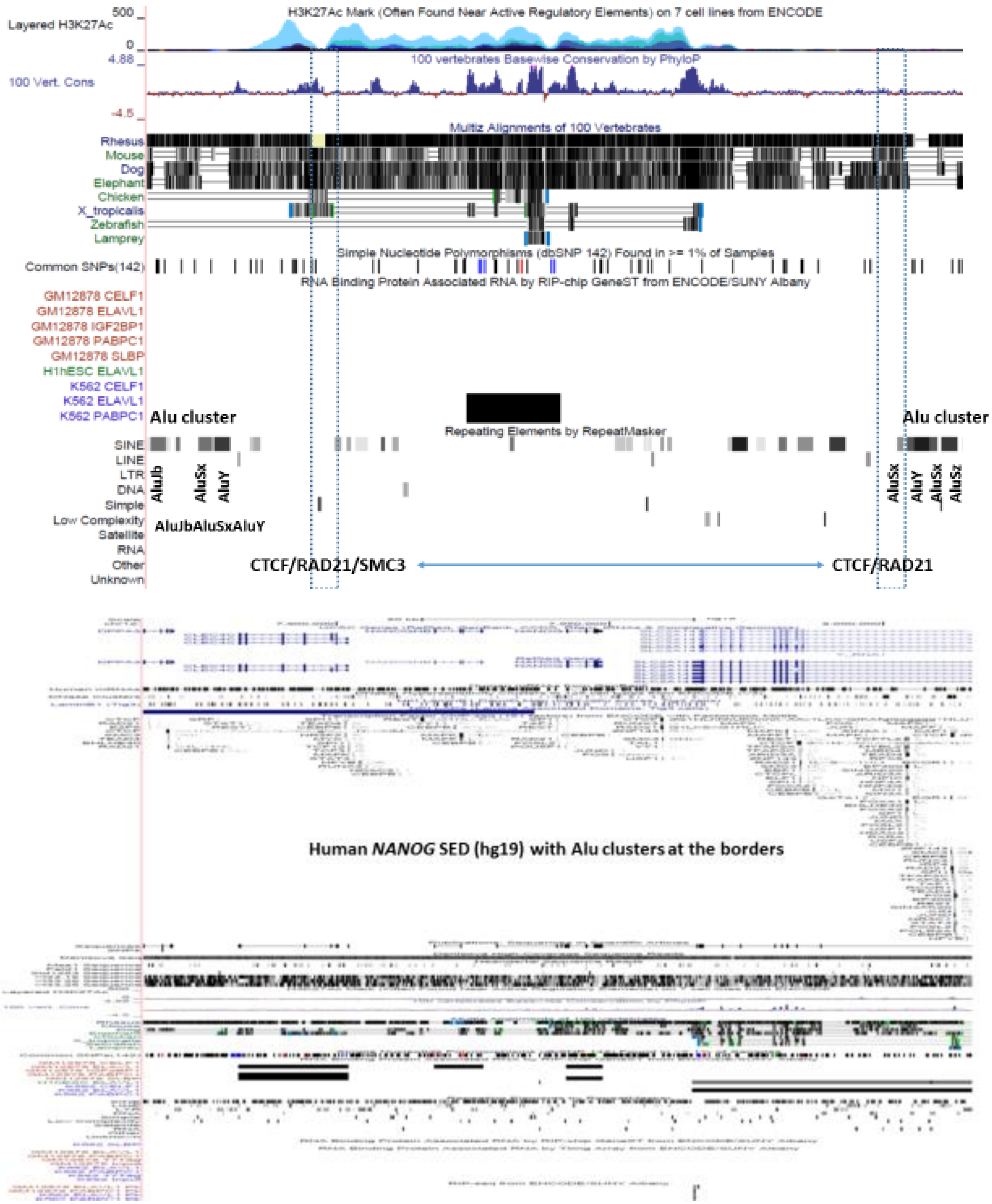

Human *NANOG* SED (hg19) with Alu clusters at the borders

Supplemental Figure S3. Related to the Table 4.
A majority of novel TFBS in the hESC genome is created by exaptation of conserved DNA sequences.

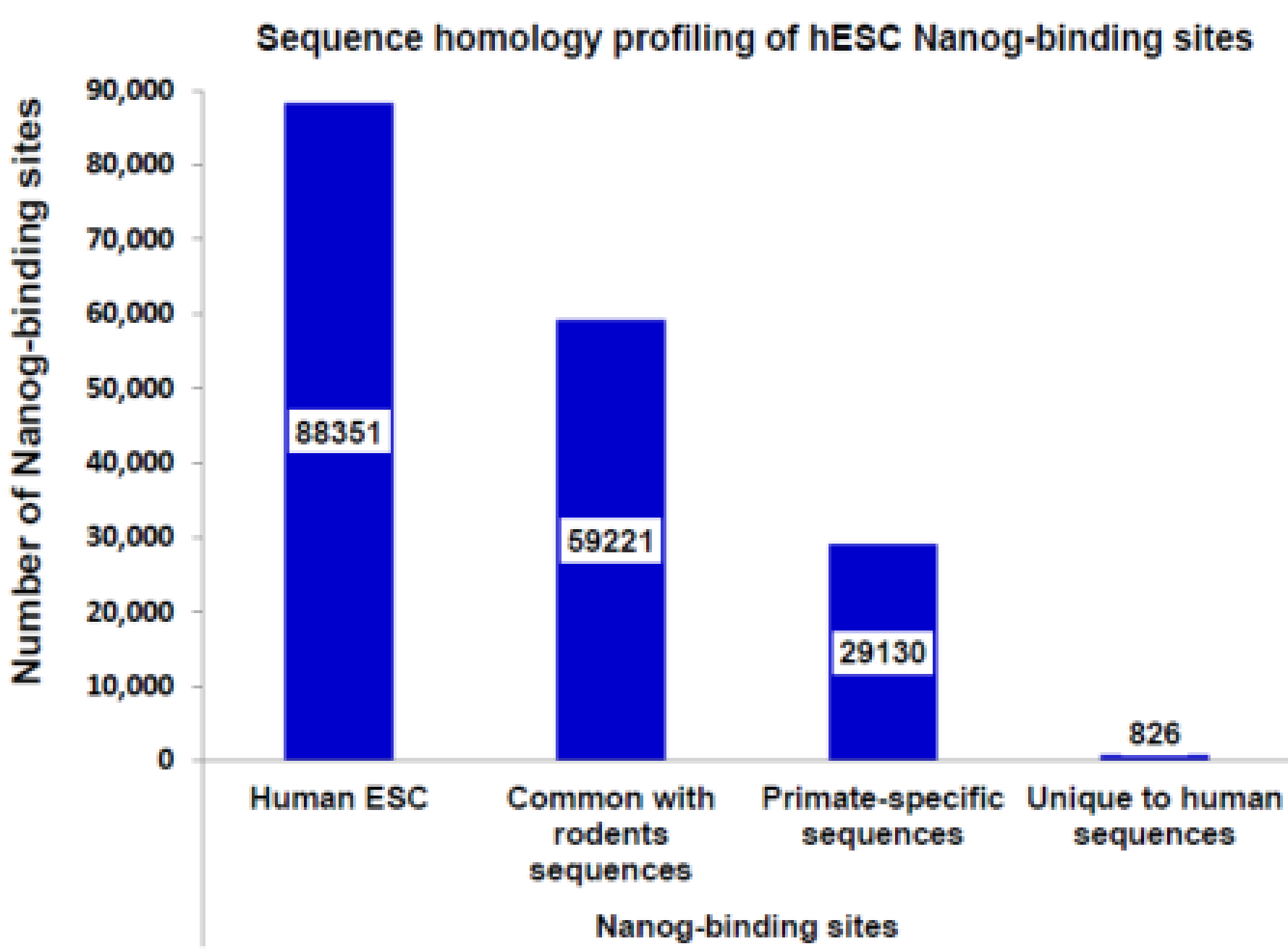

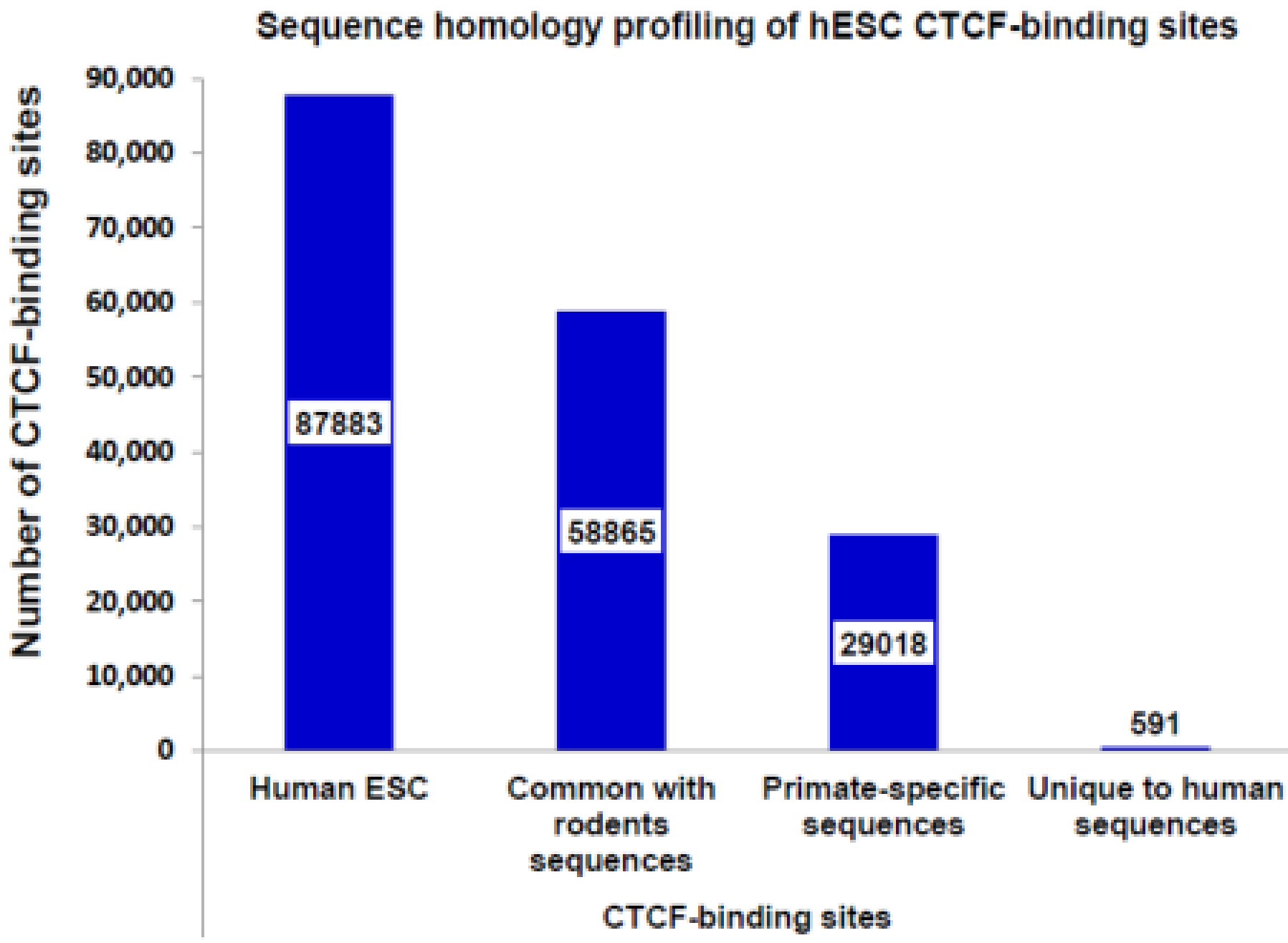

Supplemental Figure S4. Related to the Figure 6.
Additional examples of the HSGRL conservation analyses in individual genomes of 12 Modern Humans, 41,000 years old human genome from Denisova cave, and 3 Neanderthals.

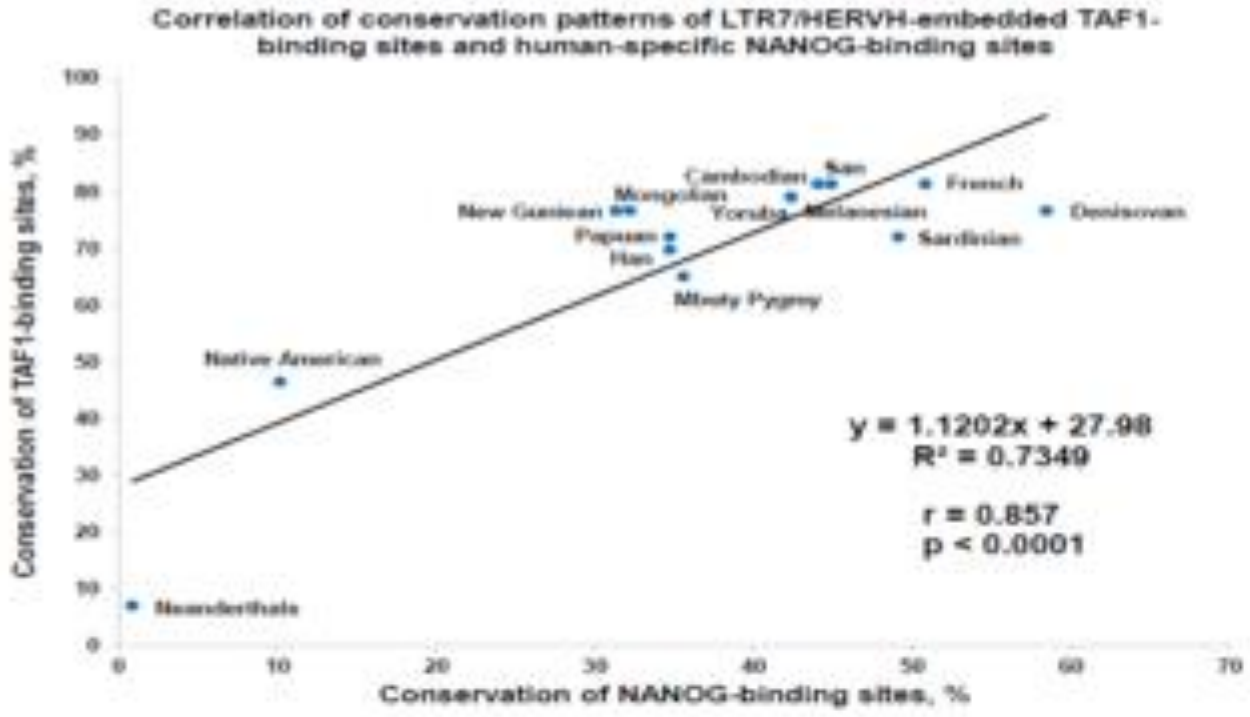
Correlation of conservation patterns of LTR7/HERVH-embedded TAF1-binding sites and human-specific NANOG-binding sites
Conservation of TAF1-binding sites, %
Conservation of NANOG-binding sites, %
y = 1.1202x + 27.98
R² = 0.7349
r = 0.857
p < 0.0001
Cambodian
San
French
Mongolian
New Guniean
Yoruba
Melanesian
Denisovan
Papuan
Han
Sardinian
Mbuty Pygmy
Native American
Neanderthals

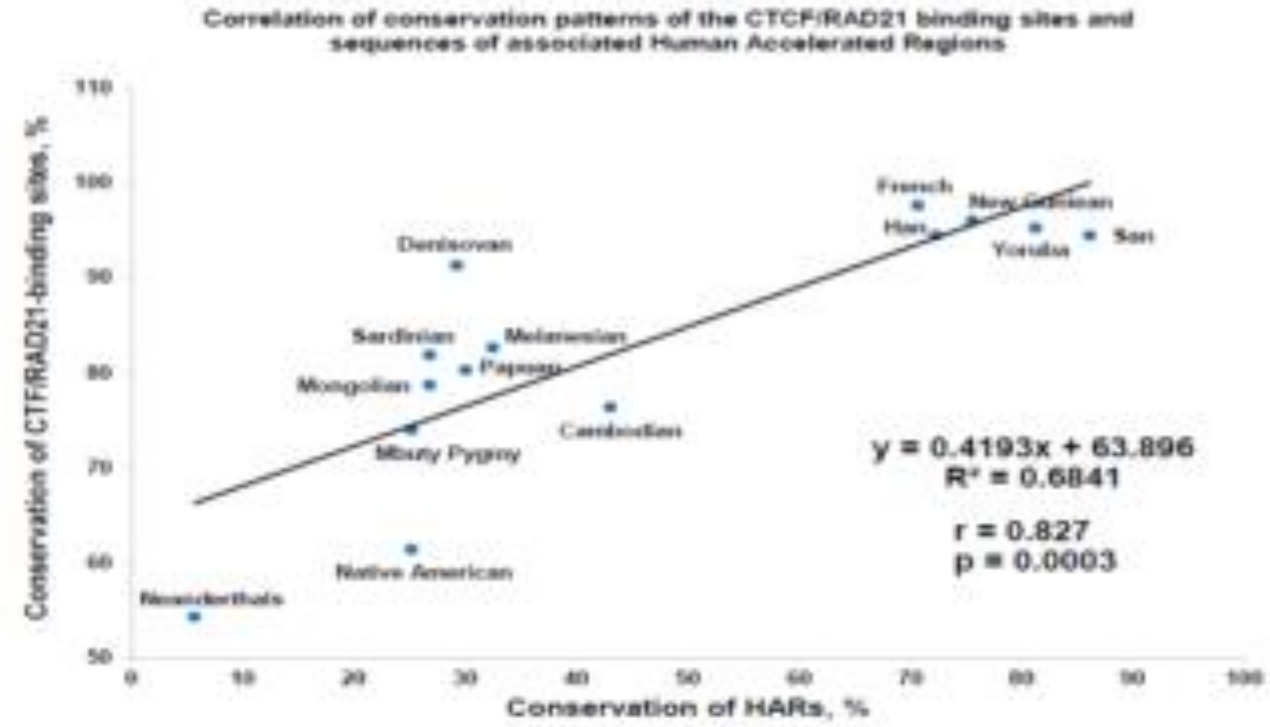
Correlation of conservation patterns of the CTCF/RAD21 binding sites and sequences of associated Human Accelerated Regions
Conservation of CTF/RAD21-binding sites, %
Conservation of HARs, %
y = 0.4193x + 63.896
R² = 0.6841
r = 0.827
p = 0.0003
French
New Guinean
Han
Yoruba
San
Denisovan
Sardinian
Melanesian
Papuan
Mongolian
Cambodian
Mbuty Pygmy
Native American
Neanderthals

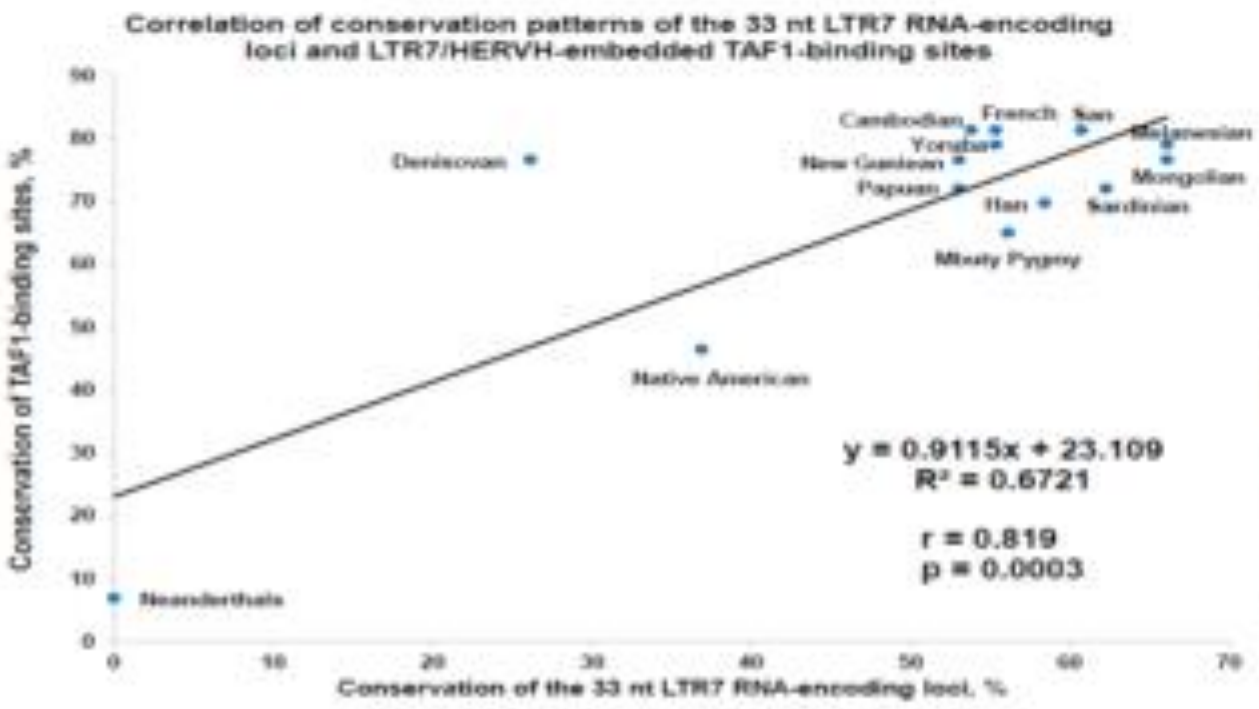
Correlation of conservation patterns of the 33 nt LTR7 RNA-encoding loci and LTR7/HERVH-embedded TAF1-binding sites
Conservation of TAF1-binding sites, %
Conservation of the 33 nt LTR7 RNA-encoding loci, %
y = 0.9115x + 23.109
R² = 0.6721
r = 0.819
p = 0.0003
Cambodian
French
San
Yoruba
Melanesian
Denisovan
New Guniean
Papuan
Han
Mongolian
Sardinian
Mbuty Pygmy
Native American
Neanderthals

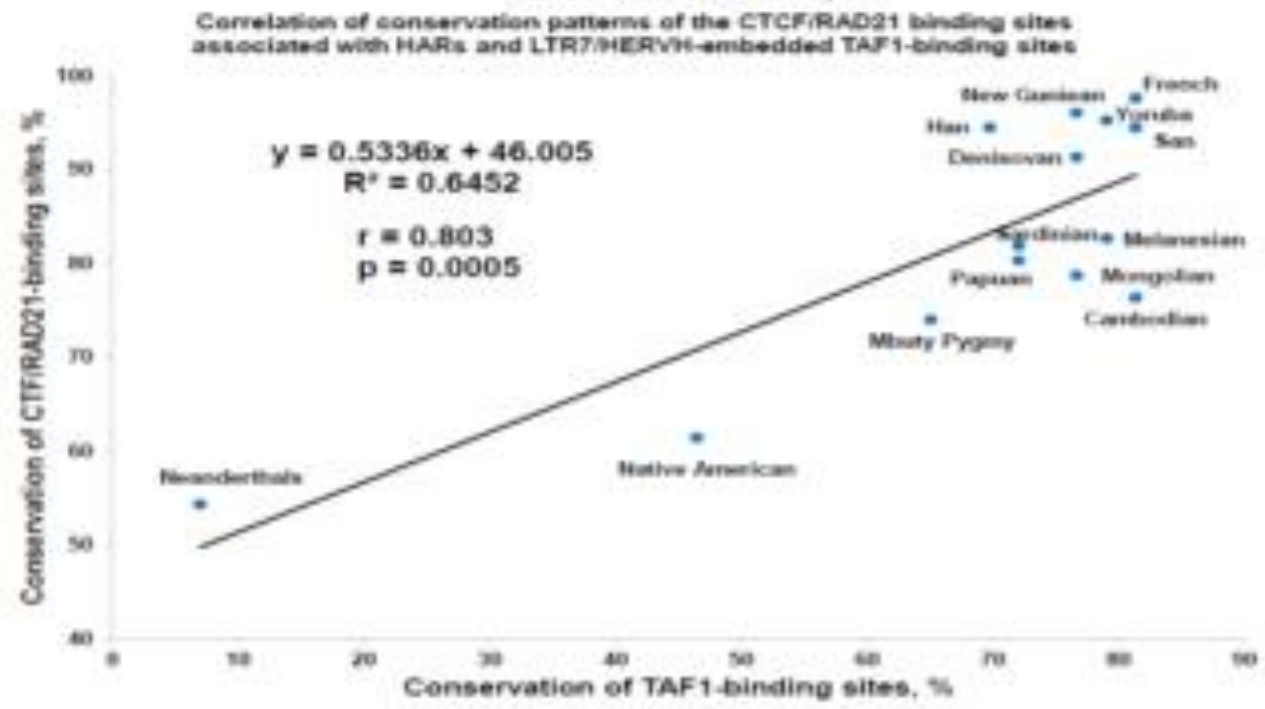
Correlation of conservation patterns of the CTCF/RAD21 binding sites associated with HARs and LTR7/HERVH-embedded TAF1-binding sites
Conservation of CTF/RAD21-binding sites, %
Conservation of TAF1-binding sites, %
y = 0.5336x + 46.005
R² = 0.6452
r = 0.803
p = 0.0005
New Guniean
French
Yoruba
Han
San
Denisovan
Sardinian
Melanesian
Papuan
Mongolian
Cambodian
Mbuty Pygmy
Native American
Neanderthals

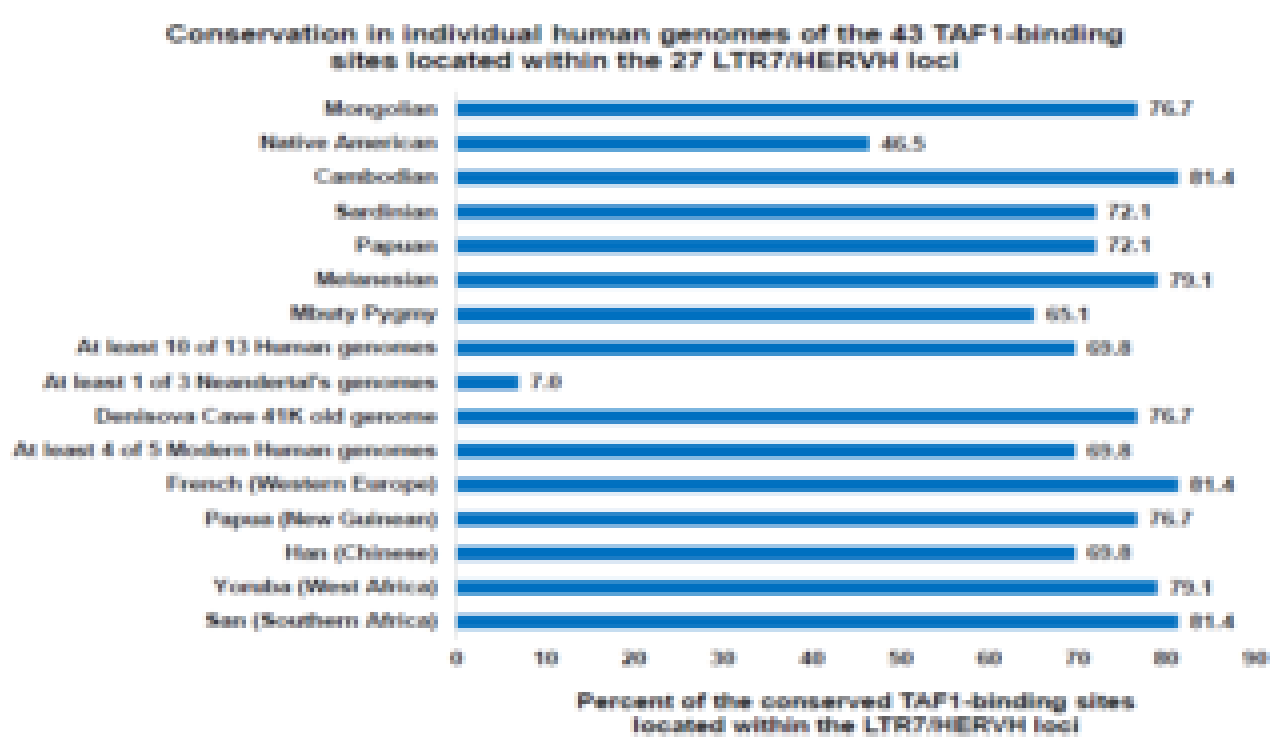
Conservation in individual human genomes of the 43 TAF1-binding sites located within the 27 LTR7/HERVH loci
Mongolian 76.7
Native American 46.5
Cambodian 81.4
Sardinian 72.1
Papuan 72.1
Melanesian 79.1
Mbuty Pygmy 65.1
At least 10 of 13 Human genomes 69.8
At least 1 of 3 Neandertal's genomes 7.0
Denisova Cave 41K old genome 76.7
At least 4 of 5 Modern Human genomes 69.8
French (Western Europe) 81.4
Papua (New Guinean) 76.7
Han (Chinese) 69.8
Yoruba (West Africa) 79.1
San (Southern Africa) 81.4
Percent of the conserved TAF1-binding sites located within the LTR7/HERVH loci

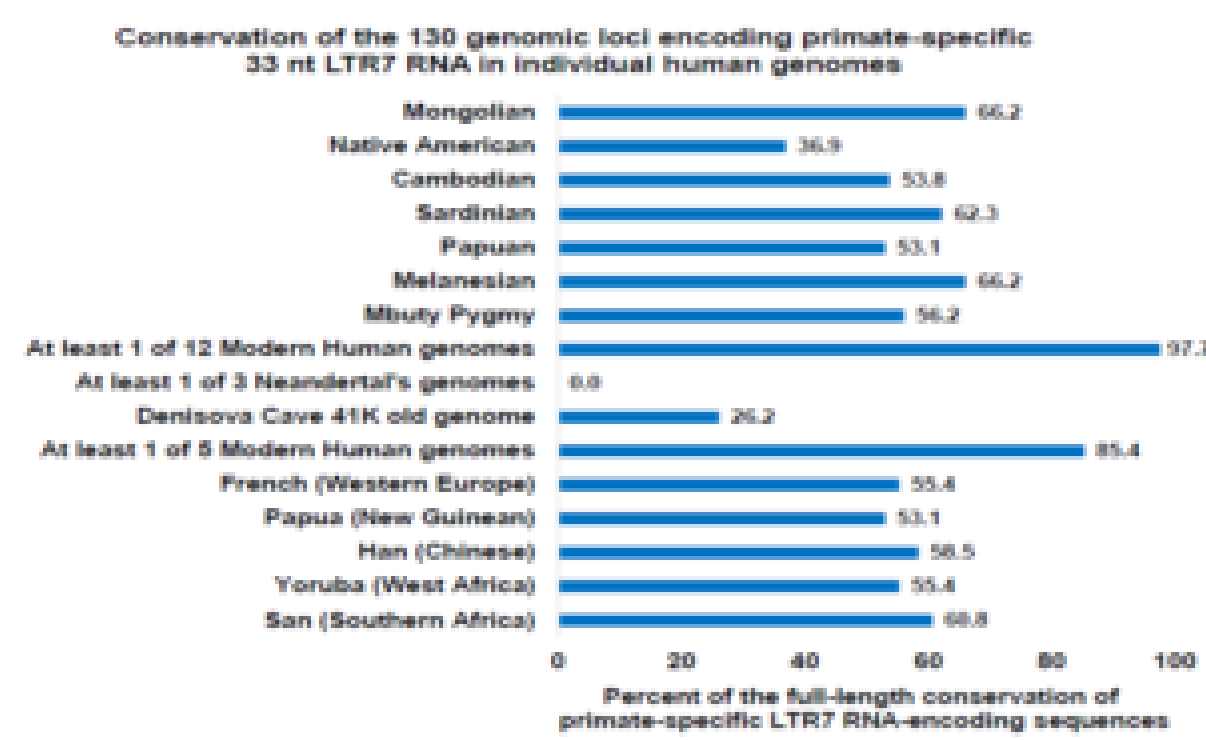
Conservation of the 130 genomic loci encoding primate-specific 33 nt LTR7 RNA in individual human genomes
Mongolian 66.2
Native American 36.9
Cambodian 53.8
Sardinian 62.3
Papuan 53.1
Melanesian 66.2
Mbuty Pygmy 56.2
At least 1 of 12 Modern Human genomes 97.7
At least 1 of 3 Neandertal's genomes 0.0
Denisova Cave 41K old genome 26.2
At least 1 of 5 Modern Human genomes 85.4
French (Western Europe) 55.4
Papua (New Guinean) 53.1
Han (Chinese) 58.5
Yoruba (West Africa) 55.4
San (Southern Africa) 60.8
Percent of the full-length conservation of primate-specific LTR7 RNA-encoding sequences

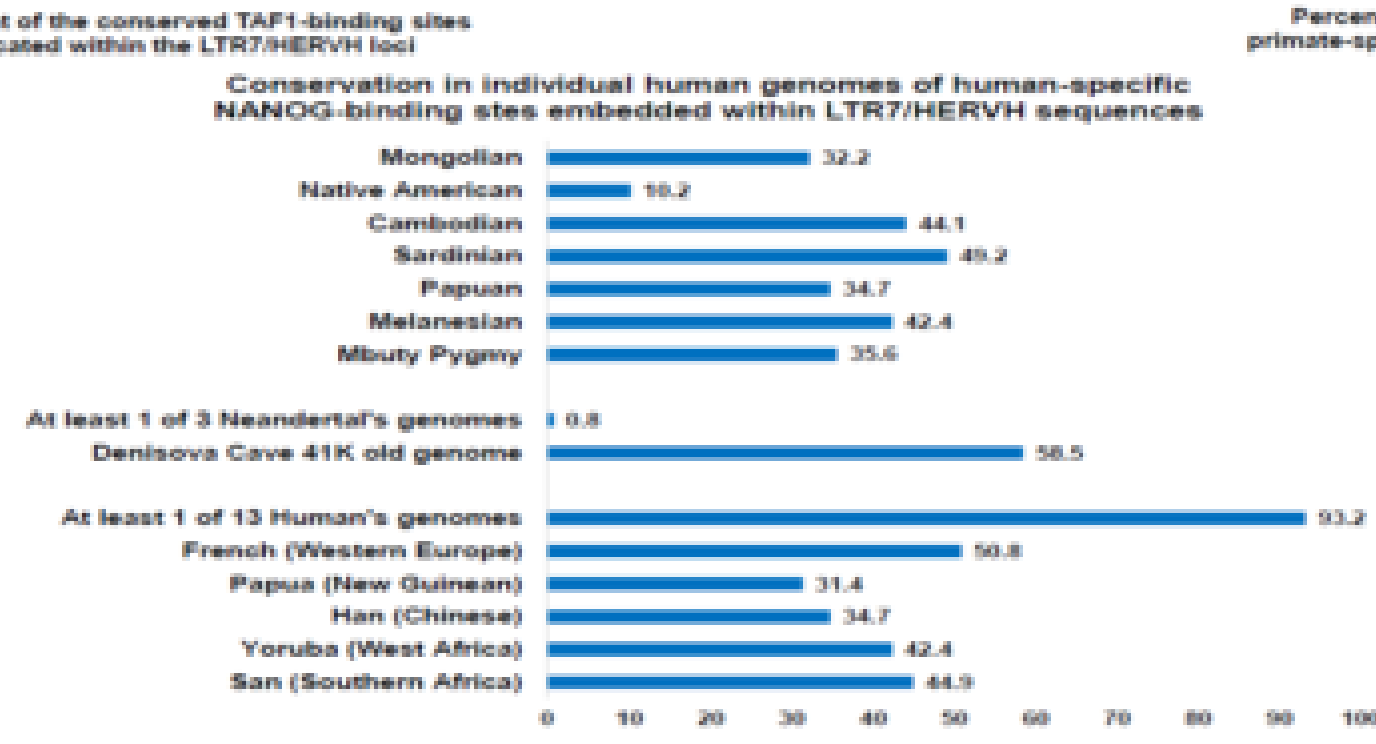
Conservation in individual human genomes of human-specific NANOG-binding stes embedded within LTR7/HERVH sequences
Mongolian 32.2
Native American 10.2
Cambodian 44.1
Sardinian 49.2
Papuan 34.7
Melanesian 42.4
Mbuty Pygmy 35.6
At least 1 of 3 Neandertal's genomes 0.8
Denisova Cave 41K old genome 58.5
At least 1 of 13 Human's genomes 93.2
French (Western Europe) 50.8
Papua (New Guinean) 31.4
Han (Chinese) 34.7
Yoruba (West Africa) 42.4
San (Southern Africa) 44.9
Conservation of human-specific NANOG-binding sites, %

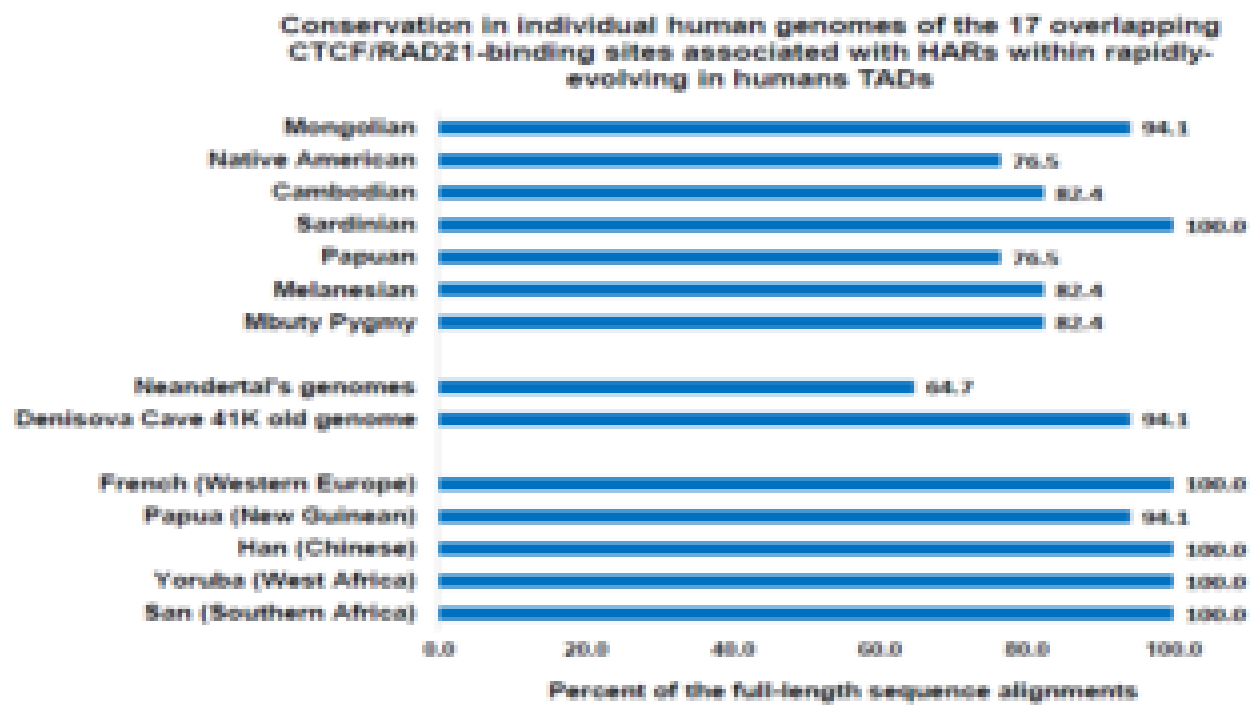
Conservation in individual human genomes of the 17 overlapping CTCF/RAD21-binding sites associated with HARs within rapidly-evolving in humans TADs
Mongolian
Native American
Cambodian
Sardinian
Papuan
Melanesian
Mbuty Pygmy
Neandertal's genomes
Denisova Cave 41K old genome
French (Western Europe)
Papua (New Guinean)
Han (Chinese)
Yoruba (West Africa)
San (Southern Africa)
94.1
76.5
82.4
100.0
76.5
82.4
82.4
64.7
94.1
100.0
94.1
100.0
100.0
100.0
0.0
20.0
40.0
60.0
80.0
100.0
Percent of the full-length sequence alignments

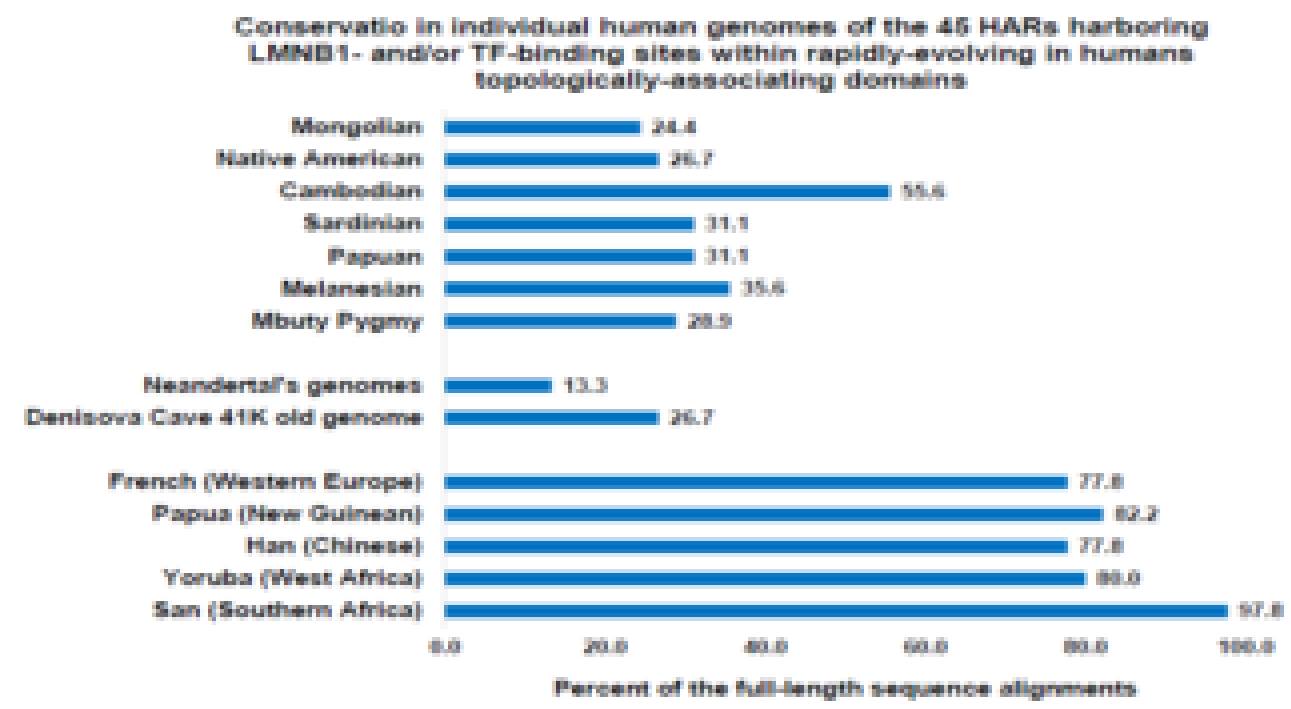
Conservatio in individual human genomes of the 45 HARs harboring LMNB1- and/or TF-binding sites within rapidly-evolving in humans topologically-associating domains
Mongolian
Native American
Cambodian
Sardinian
Papuan
Melanesian
Mbuty Pygmy
Neandertal's genomes
Denisova Cave 41K old genome
French (Western Europe)
Papua (New Guinean)
Han (Chinese)
Yoruba (West Africa)
San (Southern Africa)
24.4
26.7
55.6
31.1
31.1
35.6
28.9
13.3
26.7
77.8
82.2
77.8
80.0
97.8
0.0
20.0
40.0
60.0
80.0
100.0
Percent of the full-length sequence alignments